\documentclass[traditabstract]{aa}
\usepackage{natbib,epsfig,graphicx,ulem}
\usepackage{amsmath,amsfonts,amssymb}
\usepackage{txfonts}
\usepackage{longtable}
\usepackage{soul,color,lscape}
\usepackage[breaklinks,colorlinks,citecolor=blue]{hyperref}

\newcommand{\kpc}{{\mathrm{kpc}}}

\newcommand{\ah}{{\mathrm{h_{70}^{-1}}}}

\begin{document}

\title{The faint end of the red sequence galaxy luminosity function: unveiling
  surface brightness selection effects with the CLASH clusters.
  ~\thanks{Based on publicly available HST data acquired with ACS
    through the CLASH and COSMOS surveys. Also based on Subaru
    Suprime-Cam archive data collected at the Subaru Telescope, which
    is operated by the National Astronomical Observatory of Japan.}}

\titlerunning{The faint end of the red sequence galaxy luminosity function.}
\authorrunning{Martinet et al.}

\author{Nicolas Martinet\inst{1}, Florence Durret\inst{2}, Christophe
  Adami\inst{3}, Gregory Rudnick\inst{4}} \offprints{Nicolas Martinet,
  \email{nmartinet@astro.uni-bonn.de}}

\institute{Argelander-Institut f\"ur Astronomie, Universit\"at Bonn,
  Auf dem H\"ugel 71, D-53121 Bonn, Germany \and Sorbonne
  Universit\'es, UPMC Univ Paris 6 et CNRS, UMR 7095, Institut
  d’Astrophysique de Paris, 98 bis bd Arago, 75014 Paris, France \and
  Aix Marseille Univ, CNRS, LAM, Laboratoire d'Astrophysique de
  Marseille, Marseille, France \and Department of Physics and
  Astronomy, The University of Kansas, Malott room 1082, 1251 Wescoe
  Hall Drive, Lawrence, KS 66045, USA}

\setcounter{page}{1}

\abstract {Characterizing the evolution of the faint end of the
  cluster red sequence (RS) galaxy luminosity function (GLF) with
  redshift is a milestone in understanding galaxy evolution. However,
  the community is still divided in that respect, hesitating between
  an enrichment of the RS due to efficient quenching of blue
  galaxies from $z\sim1$ to present-day or a scenario in which the RS
  is built at a higher redshift and does not evolve
  afterwards. Recently, it has been proposed that surface brightness
  (SB) selection effects could possibly solve the literature
  disagreement, accounting for the diminishing of the RS faint
  population in ground based observations. We investigate this
  hypothesis by comparing the RS GLFs of 16 CLASH clusters computed
  independently from ground-based Subaru/Suprime-Cam V and Ip or
  Ic and space-based HST/ACS F606W and F814W images in the
  redshift range $0.187\leq z\leq0.686$. We stack individual cluster
  GLFs in two redshift ($0.187\leq z \leq0.399$ and $0.400\leq
  z\leq0.686$) and two mass ($6 \times 10^{14}M_\odot\leq M_{200} <
  10^{15}M_\odot$ and $10^{15}M_\odot\leq M_{200}$) bins, and also
  measure the evolution with the enclosing radius from 0.5~Mpc up to
  the virial radius for the Subaru large field of view
  data. Finally, we simulate the low redshift clusters at higher
    redshift to investigate SB dimming effects.

  We find similar RS GLFs for space and ground based data, with a
    difference of 0.2$\sigma$ in the faint end parameter $\alpha$ when
    stacking all clusters together and a maximum difference of
    0.9$\sigma$ in the case of the high redshift stack, demonstrating
    a weak dependence on the type of observations in the probed
    range of redshift and mass. When considering the full sample, we
    estimate $\alpha = -0.76 \pm 0.07$ and $\alpha = -0.78 \pm 0.06$
    with HST and Subaru respectively. We note a mild variation of the
    faint end between the high and low redshift subsamples at a
    1.7$\sigma$ and 2.6$\sigma$ significance.  We investigate the
      effect of SB dimming by simulating our low redshift galaxies at
      high redshift. We measure an evolution in the faint end slope of
      less than 1$\sigma$ in this case, implying that the observed
      signature is moderately larger than one would expect from SB
      dimming alone, and indicating a true evolution in the faint end
      slope.
    Finally, we find no variation with mass or radius in the probed
    range of these two parameters. We therefore conclude that
    quenching is mildly affecting cluster galaxies at $z\lesssim0.7$
    leading to a small enrichment of the RS until today, and that the
    different faint end slopes observed in the literature are probably
    due to specific cluster-to-cluster variation.}


\maketitle

\section{Introduction}
\label{sec:intro}

The study of nearby galaxy clusters led to the consensus that they
contain a rich population of red, mostly early type, galaxies. These galaxies lie
on the so-called red sequence (RS) in a color magnitude diagram, and
their galaxy luminosity function (GLF) shows a flat faint end
\citep[e.g. ][]{Gaidos97,Paolillo+01}. However, the evolution of the
faint red sequence population at higher redshift (until $z\sim1$) is
still debated. Some authors detect a strong decrease of this
population at higher redshift and optical wavelength
\citep[e.g. ][]{Smail+98,DeLucia+04,Tanaka+05,DeLucia+07,Stott+07,Gilbank+08,Rudnick+09,Vulcani+11,Martinet+15a},
highlighting an efficient quenching of the blue galaxies in
dense environments. Others find a red sequence GLF faint end constant
with redshift in the same wavelength range
\citep[e.g. ][]{Andreon06,DePropris+07,DePropris+13}. Recently,
\citet{Zenteno+16} also reported a mild evolution of the RS GLF faint
end at a $2.1\sigma$ level in the redshift range
$0.10<z<1.13$. Solving this apparent contradiction in the literature
is mandatory for understanding the evolution of galaxies, and in
particular the process that drives the quenching of the star forming
galaxies in clusters.

Interestingly, most of the studies in favor of a constant GLF faint
end are based on Hubble Space Telescope (HST) data,
e.g. \citet{Andreon06,DePropris+13}. One of the most promising effects
to solve the literature discrepancy is therefore surface brightness (SB)
selection. As proposed by \citet{DePropris+13} space based data should
detect a higher number of faint galaxies than ground based data as the darker sky in space increases the surface brightness sensitivity compared to most ground-based surveys, and
this could explain why the faint end RS population is lower in some
studies. Another explanation could be the area in which cluster
properties are measured. The HST field of view being typically much
lower than that of ground based telescopes, HST observations will only
probe the core of clusters. There could also be a dependence of the
faint end population on cluster mass, with a more efficient quenching
in the larger clusters, though there could be substantial
  cluster-to-cluster variation at a given cluster mass. In addition to possible difference between ground based and space based image sensitivity to SB, the $\propto (1+z)^4$ SB dimming with redshift could explain the
  observed RS faint end evolution without a need for additional quenching. Although cluster member selection can also contribute to the literature discrepancy, we do not study its effect in the present paper.

The goal of this paper is to shed light on the redshift evolution of
the RS faint end of cluster GLFs at optical wavelengths. To uncover SB
selection effects, we study the GLFs of clusters independently from
HST/ACS space-based data and Subaru/Suprime-Cam ground-based data. We
make use of the CLASH clusters which span the redshift range
$0.187\leq z\leq0.686$ and have images in both instruments. The
comparison of the GLF faint end derived from space and ground based
data allows to quantify the effect of SB selection that might
occur. We also simulate SB dimming to see whether it could
  explain the faint end evolution. In addition we make use of the
accurate masses derived from weak and strong lensing in
\citet{Umetsu+15} to study the variation of the faint end with cluster
mass. Finally, we compute GLFs from the cluster cores out to
  various fractions of the virial radii with the Suprime-Cam data to
investigate any dependence of the faint end on radius.

The paper is structured as follows. We first review the data we are
using in Sect.~\ref{sec:data}. We then describe the methodology to
compute individual and composite RS GLFs in Sect.~\ref{sec:glf}. We
show our results in Sect.~\ref{sec:res} and interpret them in the
discussion (Sect.~\ref{sec:discussion}). We use AB magnitudes
  throughout the paper, and assume a flat $\Lambda$CDM cosmology
  with $\Omega_{\rm M}=0.3$ and $h=0.7$.

\section{Data}
\label{sec:data}

\subsection{Clusters: CLASH}
\label{subsec:clash}

We study a sub-sample of 16 clusters from the Cluster Lensing And
Supernova survey with Hubble (CLASH) sample \citep{Postman+12} in the
redshift range $0.187\leq z\leq0.686$. We first select the 18 clusters having
both HST and Subaru data in the I and V bands to be able to select red
sequence galaxies in a color-magnitude diagram independently from
space and ground based data. We have to discard two clusters from this
sample: MACSJ0647 because of a large star halo that plagues the
cluster region, and MACSJ1931 because it is crowded with bright stars.

We make use of the reduced HST Advanced Camera for Survey (ACS)
\citep{Ford+03} and Subaru Suprime-Cam \citep{Miyazaki+02} images made
public by the CLASH collaboration, and refer the reader to the CLASH
overview paper \citep{Postman+12} for details on the reduction
process.

The coordinates and redshifts of the 16 retained clusters are
displayed in Table~\ref{tab:data}, along with the available images and
their PSF FWHM in HST/ACS F814W and F606W bands, and Subaru/Suprime-Cam
Ic or Ip and V bands. When both Ic and Ip filters are available, we use
Ic because most of the sample is imaged in Ic and it is closer to the
F814W band, decreasing the magnitude shift applied when
homogenizing the magnitudes to the HST/ACS filters. The positions
correspond to the X-ray centers except for MACSJ2129 for which the center is
derived from the optical image. The seeing for the Subaru images is
taken from the CLASH data web-page
(\url{https://archive.stsci.edu/prepds/clash/}), and is computed using
PSFEx \citep{Bertin11} for the HST images. We note that the HST PSF varies around its nominal value of $\sim0.1\arcsec$ in the optical due to single image estimates with low star density. These variations are acceptable given that we do not perform any weak lensing measurement on these data. When available, we also
display the $r_{200}$ critical radius and the ${\rm M}_{200}$ total mass
computed from joint weak and strong lensing, and magnification by
\citet{Umetsu+15}. Finally, we show the maximum usable radius for the comparison between HST and Subaru GLFs, given the HST limited field of view.

\begin{table*}
  \caption{Studied clusters from the CLASH sample. The different columns are: \#1: cluster ID, \#2: right ascension, \#3: declination, \#4: redshift, \#5: FWHM of the HST/ACS F814W image PSF in arcseconds, \#6: FWHM of the HST/ACS F606W image PSF in arcseconds, \#7: FWHM of the Subaru/Suprime-Cam Ic and Ip image PSFs in arcseconds, \#8: FWHM of the Subaru/Suprime-Cam V image PSF in arcseconds, \#9: maximum cluster radius that fits in the HST field-of-view , \#10: radius at which the cluster mass density is 200 times the critical mass density, as computed from lensing \citep{Umetsu+15}, \#11: cluster total mass, as computed from lensing \citep{Umetsu+15}.}  \centering
\begin{tabular}{lcccccccccc}
  \hline
  \hline 
  Cluster & RA & DEC & z & $\epsilon_{\rm F814W}$ & $\epsilon_{\rm F606W}$ & $\epsilon_{\rm Ic}$/$\epsilon_{\rm Ip}$ & $\epsilon_{\rm V}$ & $r_{\rm max}$ & $r_{200}$ & ${\rm M}_{200}$\\
   &  &  &  & $\arcsec$ & $\arcsec$ & $\arcsec$/$\arcsec$ & $\arcsec$ & $\kpc.\ah$ & $\kpc.\ah$ & $10^{14}~{\rm M}_\odot.\ah$\\
 \hline 

Abell 209      &  01:31:52.57	 & -13:36:38.8    & 0.206     & 0.06	 & 0.17	 & -/0.66	   & 0.73  & 382 &  2268  &  15.40 $\pm$ 3.42 \\
Abell 383      &  02:48:03.36	 & -03:31:44.7    & 0.187     & 0.09	 & 0.06	 & 0.86/0.57       & 0.63  & 379 &  1829  &  7.98  $\pm$ 2.66      \\
MACSJ0329-02   &  03:29:41.68	 & -02:11:47.7    & 0.450     & 0.25	 & 0.15	 & 0.90/-	   & 0.55  & 643 &  1742  &  8.65  $\pm$ 1.97 \\
MACSJ0429-02   &  04:29:36.10	 & -02:53:08.0    & 0.399     & 0.14	 & 0.09	 & 1.28/-	   & 1.14  & 610 &  1840  &  9.76  $\pm$ 3.50 \\
MACSJ0717+37   &  07:17:31.65    & +37:45:18.5    & 0.548     & 0.14	 & 0.15	 & -/0.96	   & 0.69  & 807 &  2358  &  26.77 $\pm$ 5.36 \\
MACSJ0744+39   &  07:44:52.80    & +39:27:24.4    & 0.686     & 0.08	 & 0.06	 & 0.82/0.87       & 0.71  & 859 &  2030  &  18.03 $\pm$ 4.96      \\
Abell 611      &  08:00:56.83    & +36:03:24.1    & 0.288     & 0.10	 & 0.08	 & 0.76/0.81       & 0.85  & 389 &  2223  &  15.76 $\pm$ 4.49      \\
MACSJ1115+01   &  11:15:52.05    & +01:29:56.6    & 0.352     & 0.09	 & 0.06	 & 0.96/-	   & 0.95  & 562 &  2250  &  16.66 $\pm$ 3.85 \\
MACSJ1206-08   &  12:06:12.28	 & -08:48:02.4    & 0.440     & 0.05	 & 0.16	 & 0.71/-	   & 0.95  & 643 &  2220  &  18.17 $\pm$ 4.23 \\
RXJ1347-1145   &  13:47:30.59	 & -11:45:10.1    & 0.451     & 0.11	 & 0.09	 & 1.14/-	   & 0.75  & 677 &  2720  &  34.25 $\pm$ 8.78 \\
MACSJ1423+24   &  14:23:47.76    & +24:04:40.5    & 0.545     & 0.08	 & 0.13	 & 0.86/-	   & 0.96  & 751 &  -     &        -          \\
RXJ1532.9+3021 &  15:32:53.78    & +30:20:58.7    & 0.345     & 0.08	 & 0.14	 & 1.11/-	   & 0.71  & 553 &  1508  &  5.98  $\pm$ 2.32 \\
MACSJ1720+35   &  17:20:16.95    & +35:36:23.6    & 0.391     & 0.09	 & 0.12	 & 1.04/-	   & 0.82  & 559 &  2091  &  14.50 $\pm$ 4.30 \\
MACSJ2129-07   &  21:29:26.06    & -07:41:28.8    & 0.570     & 0.05	 & 0.09	 & 0.55/-	   & 0.72  & 800 &  -     &        -	       \\
RXJ2129+0005   &  21:29:39.94    & +00:05:18.8    & 0.234     & 0.11	 & 0.06	 & -/1.00	   & 0.71  & 421 &  1680  &  6.14  $\pm$ 1.79 \\
MS 2137.3-2353 &  21:40:15.18	 & -23:39:40.7    & 0.313     & 0.12	 & 0.09	 & 1.20/-	   & 1.15  & 516 &  2160  &  13.56 $\pm$ 5.27 \\

  \hline 
  \hline
\end{tabular}
\label{tab:data}
\end{table*}

\subsection{Field: COSMOS}
\label{subsec:cosmos}

The field galaxies are measured from the COSMOS survey. We use the
HST/ACS images in the F814W and F606W filters and the
Subaru/Suprime-Cam images in the Ip and V bands, as reduced by the
3D-HST team (\url{http://3dhst.research.yale.edu/Home.html}). Details
on the image reduction can be found in \citet{Brammer+12,Skelton+14}
for the HST data and in \citet{Taniguchi+07, Capak+07} for Subaru. We
use the same telescopes and instruments as those of the cluster images
to avoid any contamination from different SB
selections. The final area of this catalog after masking is
0.0468~deg$^2$, which is more than ten times larger than any of the
cluster fields.

\section{Galaxy luminosity functions: method}
\label{sec:glf}

This section describes the methods used to build the cluster GLFs, and to
analyze them. In a nutshell, we detect objects with SExtractor,
separate stars from galaxies in a SB-magnitude
diagram, select cluster RS galaxies in a color-magnitude diagram,
compute rest-frame magnitudes using mean k-correction values derived with
LePhare, convert all magnitudes to the F814W and F606W filters, remove
COSMOS background galaxies and compute the GLFs normalized to a
1~deg$^2$ area, using Poisson error counts. Each step is done
independently on the ground Subaru and on the space HST data, so that
we can safely compare both GLFs. The GLFs are then fitted with a
Schechter function \citep{Schechter76}, to the completeness magnitude limit computed for
every image, and stacked with the Colless method \citep{Colless89}. The paragraphs below
give the details of the analysis.

\subsection{Detecting objects}
\label{subsec:detection}

While it is tempting to use the higher resolution HST data to detect
objects and then measure the magnitudes in the Subaru data at the
object positions, we refrain from doing so because we do not want to
affect the faint galaxy selection. Indeed, if we want to show evidence
of any SB selection effect, we need to detect objects
independently in the HST and Subaru data. For the same reasons, we do
not want to use already available catalogs, to make sure the
detections are done separately, and also to be able to use the
detection configuration when estimating the completeness of the images
(see Sect.~\ref{subsec:comp}).

Objects are detected using SExtractor \citep{Bertin+96} in the F814W
(resp. Ic/Ip) image for HST (resp. Subaru). Object properties are
then measured at the detected locations in double image mode in the
F606W (resp. V) band. Instead of using the same detection parameters
for HST and Subaru images, we adapt the configuration files allowing
to recover all the objects while avoiding spurious detections. This approach is closer to what is found in the literature, since one always tries to get as much information as possible from one's dataset. We also
recall that the resolutions of the two cameras, and the PSFs of each instrument, are different, requiring
different detection parameters: the ACS images have a pixel size of 0.03$\arcsec$ and a mean PSF FWHM of 0.10$\arcsec$, and the
Subaru images a pixel size of 0.2$\arcsec$ and a mean PSF FWHM of 0.87$\arcsec$. Hence, we detect objects in HST images with a minimum
 area (DETECT\_MINAREA) of 5 pixels and a minimum threshold
(DETECT\_THRESH) of 3 times the background level, and in Subaru images
with a DETECT\_MINAREA of 3 pixels and a DETECT\_THRESH of 1.5.
These values are also different
for the COSMOS data because they have been re-binned to a pixel scale
of 0.06'' for both HST and Subaru. The DETECT\_MINAREA and
DETECT\_THRESH keyword are set to 3 pixels and 1.5$\sigma$ for the
HST COSMOS images and to 5 pixels and 3$\sigma$ for the Subaru COSMOS
images. We check on every image that all objects that can be visually
identified are detected and that no spurious detections are
included after masking. This last step is done by detecting objects in the inverted image (multiplied by -1) with the same detection configuration. These objects therefore correspond to spurious detections only. We find no false detection below the completeness magnitude limit after masking. We adopt a relatively aggressive deblending strategy
(DEBLEND\_NTHRESH of 32 and DEBLEND\_MINCONT of 0.002) to be able to
detect faint cluster galaxies that could be masked by larger foreground
 objects.

We mask the areas which could lead to spurious detections: image edges, CCD inter-chips in the case of the ACS data, and bright saturated stars. These masks are
applied to every image whether it has been acquired with HST or Subaru
because for each cluster we want to study exactly the  same region with
both cameras.

Magnitudes are measured via the MAG\_AUTO algorithm implemented in SExtractor, and
are corrected for Milky Way dust extinction using maps from
\citet{Schlegel+98}. This correction factor is applied as a magnitude
shift directly in the zero point of each image. We also set the
  minimum aperture radius to 5 pixels in SExtractor so that it is
  larger than the PSF, avoiding the computations of aperture corrections
  for the MAG\_AUTO measurement when the minimum Kron radius is
  smaller than the PSF \citep[e.g. ][]{Rudnick+09}. We check this last
  point by visualizing  with DS9 the apertures in which the magnitudes are measured on the images, and find that they encompass the full galaxies up to the confusion
  with background noise. This last step is done by comparing pixel values on aperture edges with background estimates from SExtractor.

\begin{figure}[h!]
\centering
\includegraphics[angle=270,width=9.cm]{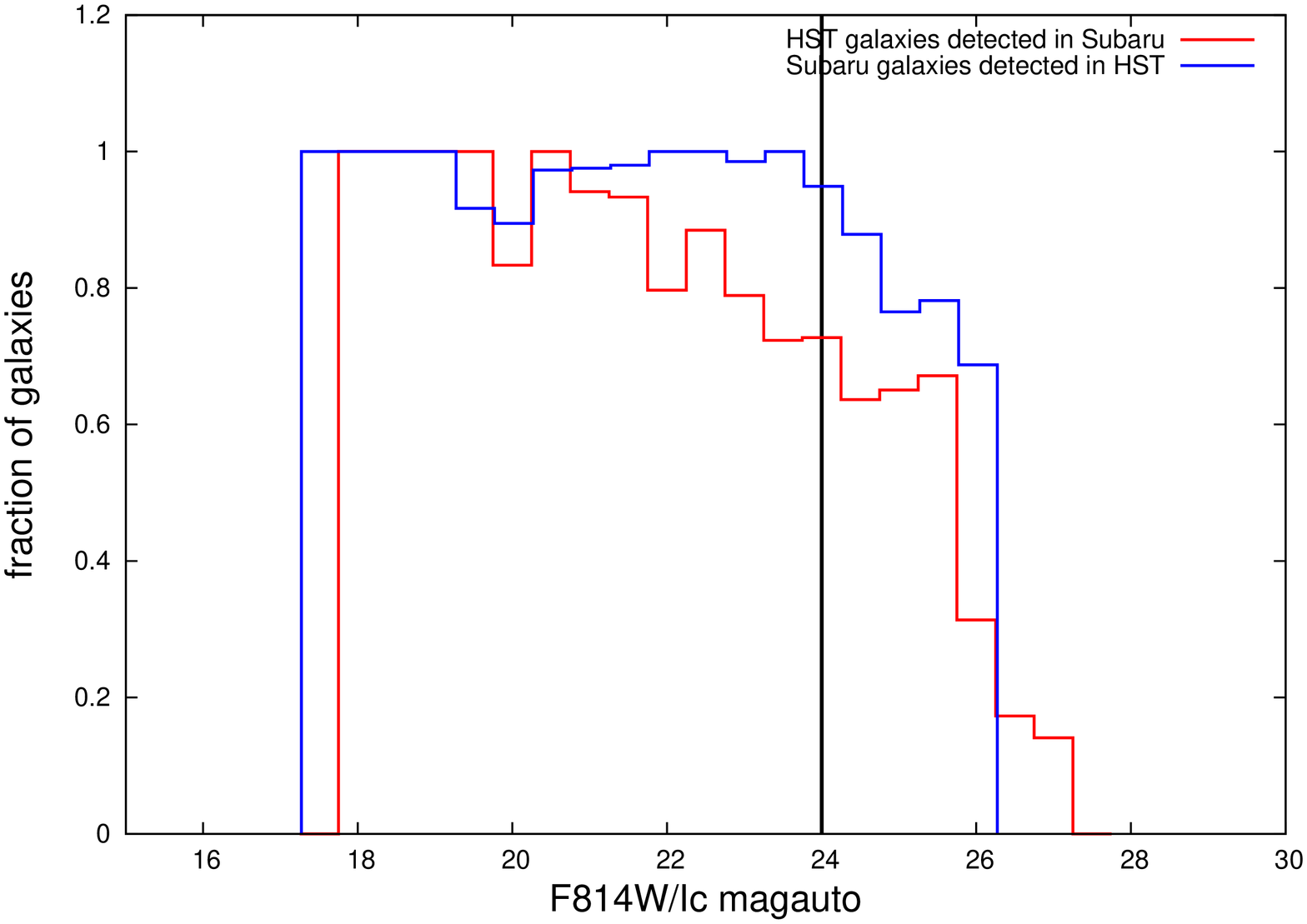}
\includegraphics[angle=270,width=9.cm]{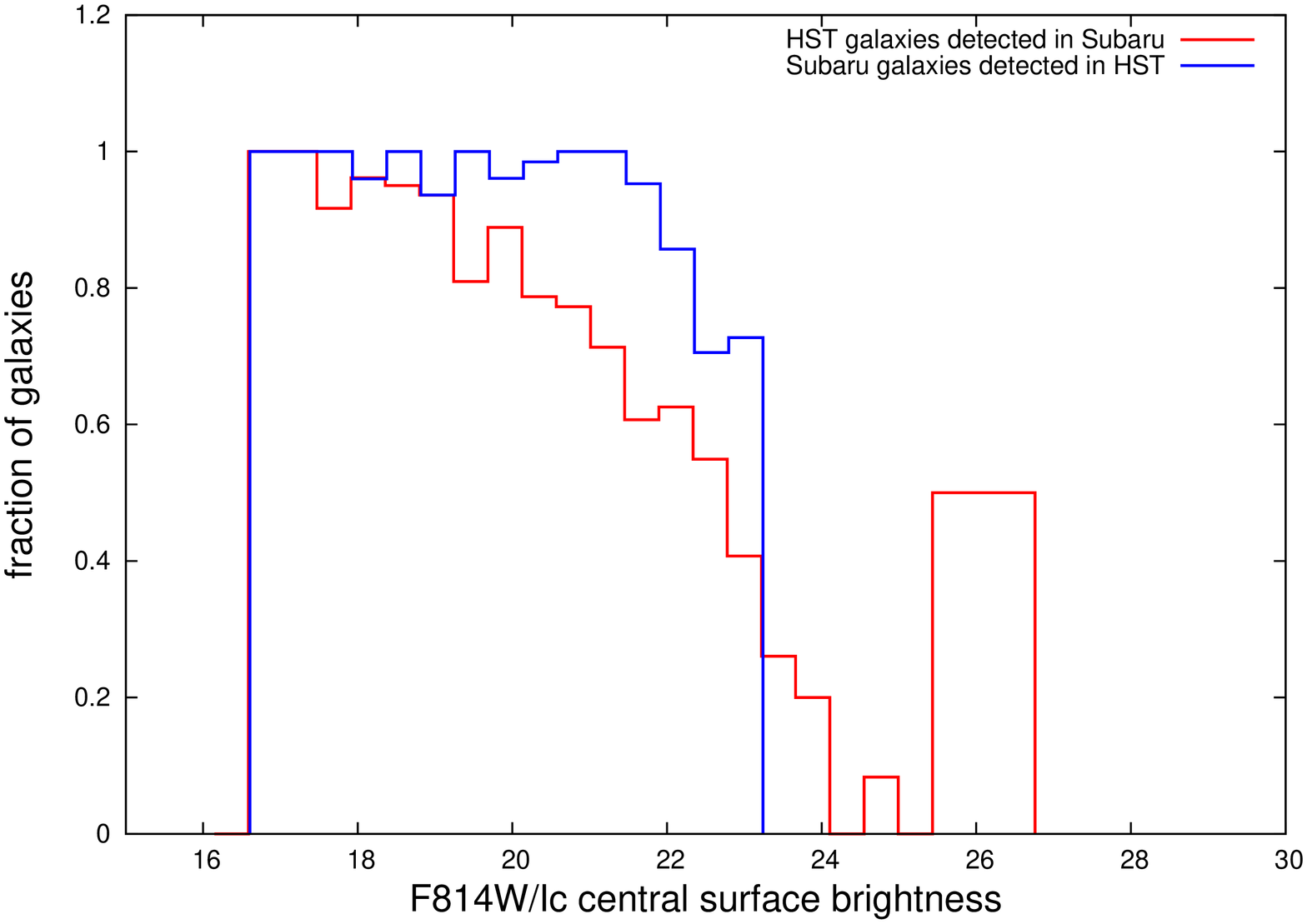}
\caption{Fraction of galaxies detected in one instrument that are redetected in the other one, as a function of magnitude ({\it top}) and central SB ({\it bottom}), for cluster RXJ1347. The red histogram corresponds to the fraction of HST galaxies redetected in the Subaru image and the blue one to the fraction of Subaru galaxies that are redetected in the HST image. The black vertical line indicates the 90\% Subaru completeness magnitude limit.}
\label{fig:redetect}
\end{figure}

We quantify the difference between the HST and Subaru detections by computing the fraction of galaxies detected in one instrument that is also detected in the other one. We first cross match the catalogs, with a nearest neighbor approach and a maximum distance criterion of $1\arcsec$, which roughly corresponds to the size of the Subaru PSF. We then measure the fraction of redetected objects, in bins of magnitude, and central SB, and show them for cluster RXJ1347 in Fig.~\ref{fig:redetect}. The central SB is estimated as the magnitude in a $0.6\arcsec$ radius centered on each galaxy, and divided by the aperture area. We see that almost all galaxies detected with Subaru are also detected in HST, and that for magnitudes fainter than 21, Subaru starts to miss galaxies compared to HST, with about 30\% of galaxies missed at the Subaru 90\% completeness limit. The effect is even clearer when plotting the histograms as a function of SB. We also note a small drop at magnitude $I=20$, although not significant because corresponding only to one or two galaxies in a bin populated by about 10 galaxies. These two plots show that there is a clear difference due to the SB selection effects between Subaru and HST, but we note that this effect also applies to field galaxies.

\subsection{Selecting red sequence galaxies}
\label{subsec:rs}

\begin{figure}[h!]
\centering
\includegraphics[angle=270,width=9.cm]{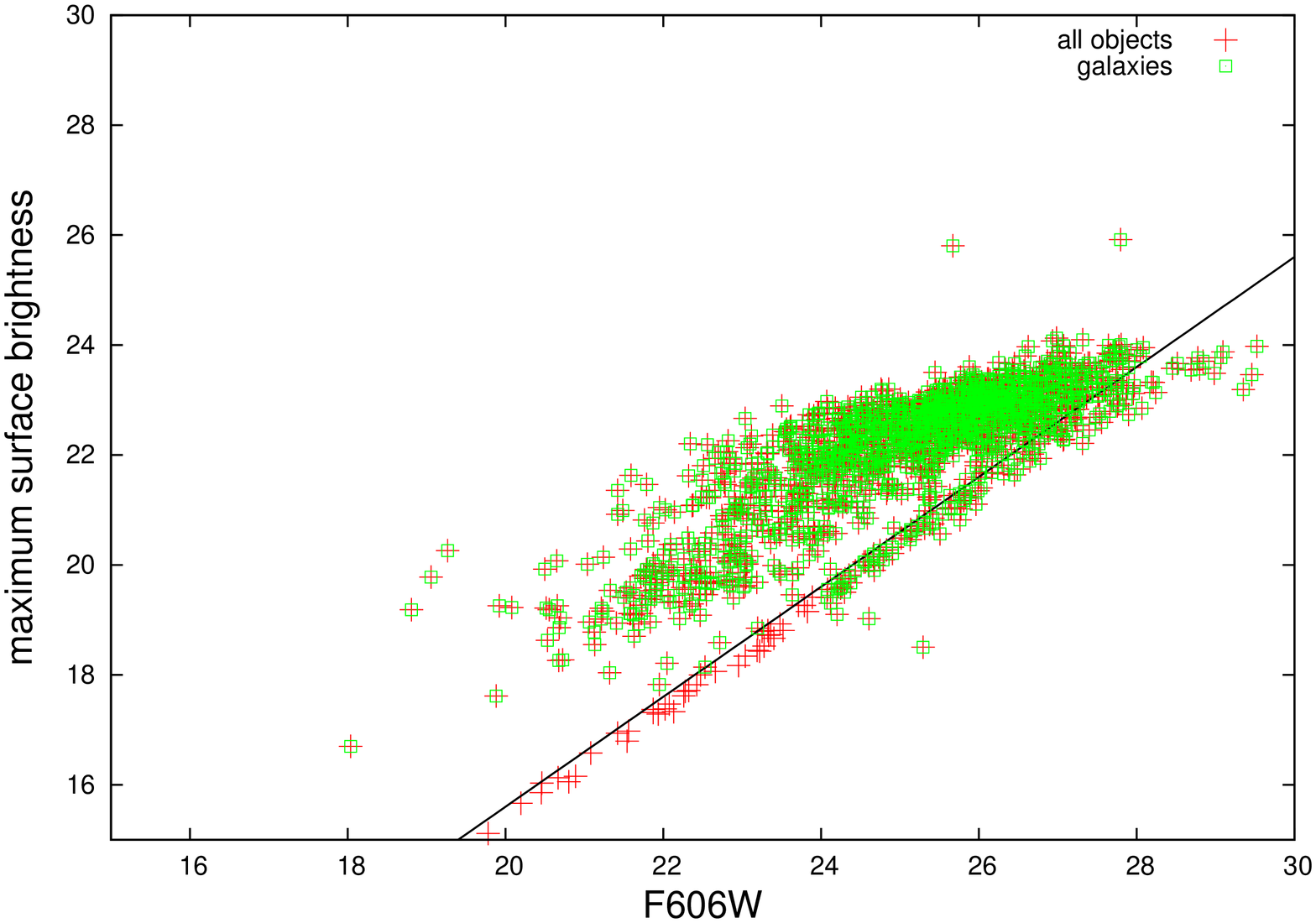}
\includegraphics[angle=270,width=9.cm]{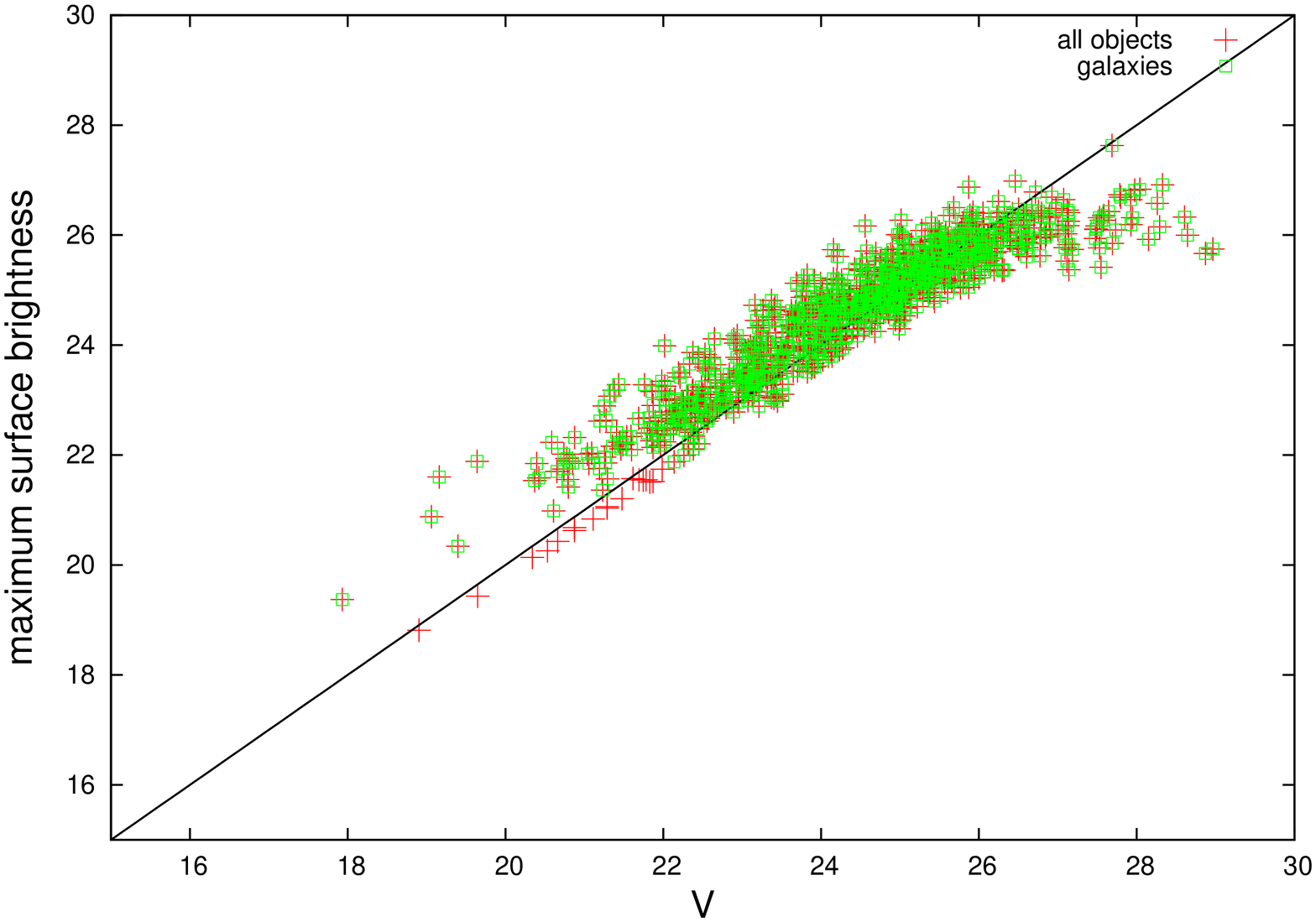}
\caption{Maximum SB versus magnitude diagrams for RXJ1347, based
    on the HST F606W data ({\it top}) and on the Subaru V data ({\it
      bottom}). Red crosses correspond to all objects before the
    galaxy-star separation, and green crosses to galaxies. Black lines
    represent the cuts applied to discard stars up to a limiting
    magnitude of F606W=24 for HST and V=22 for Subaru.}
\label{fig:starsep}
\end{figure}

Galaxies are separated from stars in a maximum SB versus magnitude
diagram. Point like sources have their maximum SB
proportional to their magnitude and can thus be isolated in this
diagram up to a certain magnitude. This separation is done
independently for the HST and Subaru catalogs, in the filter where the star sequence is best visualized. We discard stars up to F814W=23 for HST and Ic/Ip=21 for Subaru in the I band, and up to F606W=24 and V=22 when the selection is done in the V band. Above these magnitudes it
becomes difficult to make a distinction between stars and small galaxies. The
magnitude limit is higher for HST because of the smaller PSF
which allows for a better separation. We note from the Besan\c con model
\citep{Robin+03} of the Milky Way star distribution, that the number of
stars above i=21 becomes quite low compared to the observed number of
galaxies, so the remaining faint stars should not significantly
affect the GLF faint end. In addition, these stars should be bluer
than cluster galaxies in the studied redshift range, and are very
unlikely to be selected in the RS. As an illustration of the star galaxy separation we show in Fig.~\ref{fig:starsep} the diagrams of maximum SB versus magnitude with the star cuts overplotted for cluster RXJ1347.

We then cut the catalog to a circular area centered on the
cluster. The radius of this disk is set to the maximum value such that
the cluster area is fully included in the HST/ACS field of view (see Table~\ref{tab:data}). This
physical radius is then smaller for low redshift clusters. This choice
is made to use all possible data, and because we found that the
stacked GLF does not depend on the external radius. See
Sect.~\ref{subsec:rad} in which we make use of the large field of view of the
Subaru/Suprime-Cam images to compute stacked GLFs in various
radii.

\begin{figure}[h!]
\centering
\includegraphics[angle=270,width=9.cm]{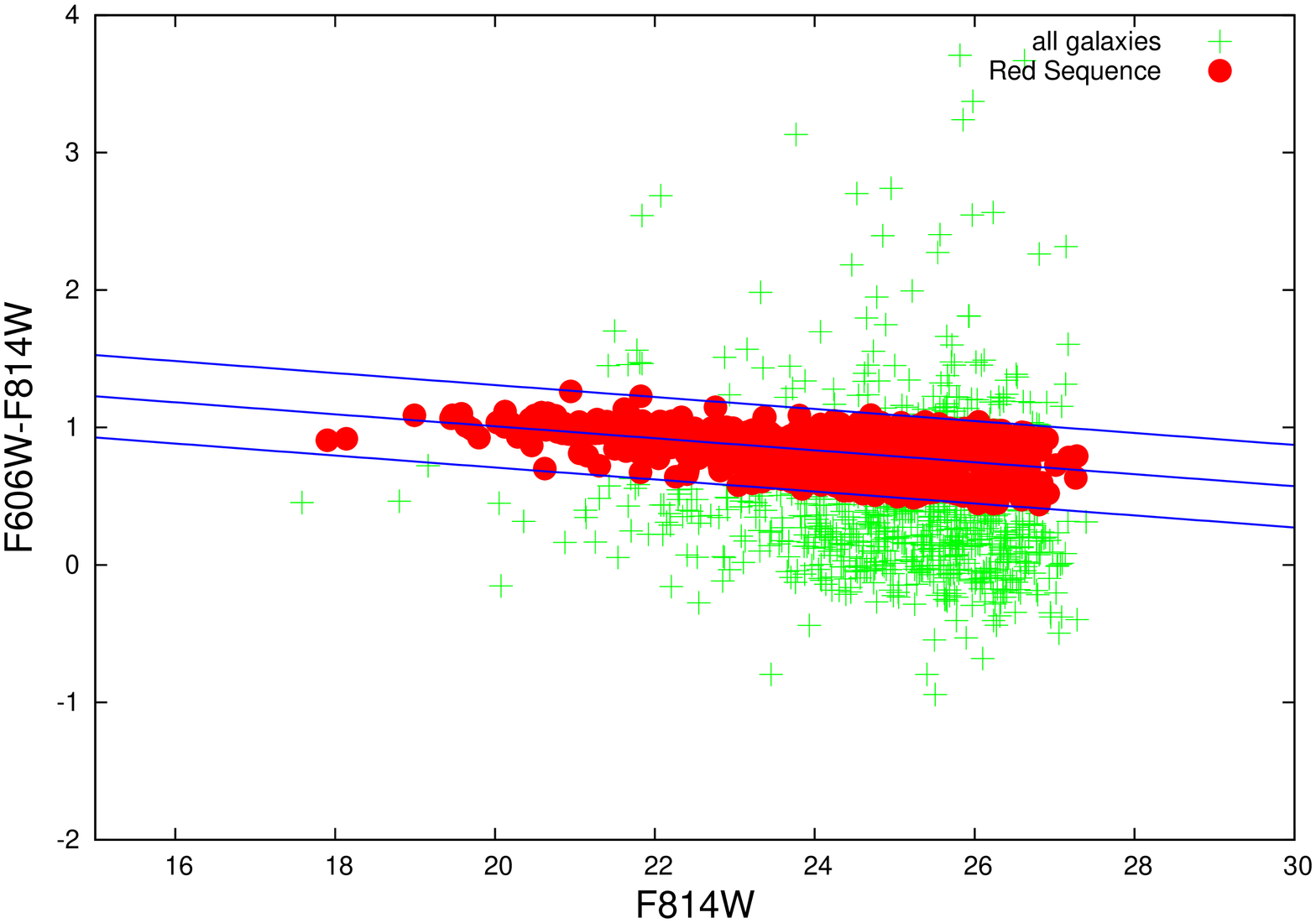}
\includegraphics[angle=270,width=9.cm]{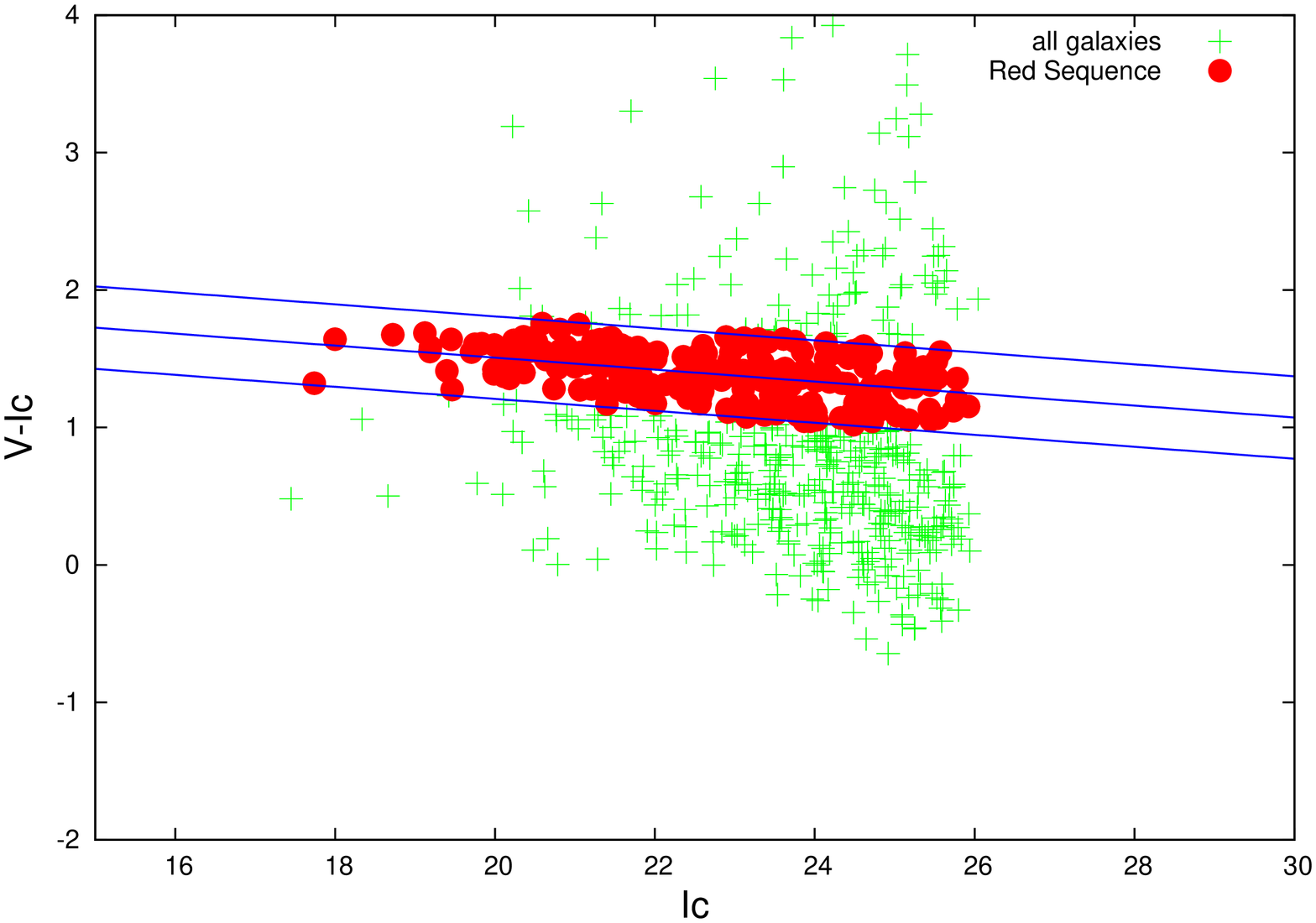}
\includegraphics[angle=270,width=9.cm]{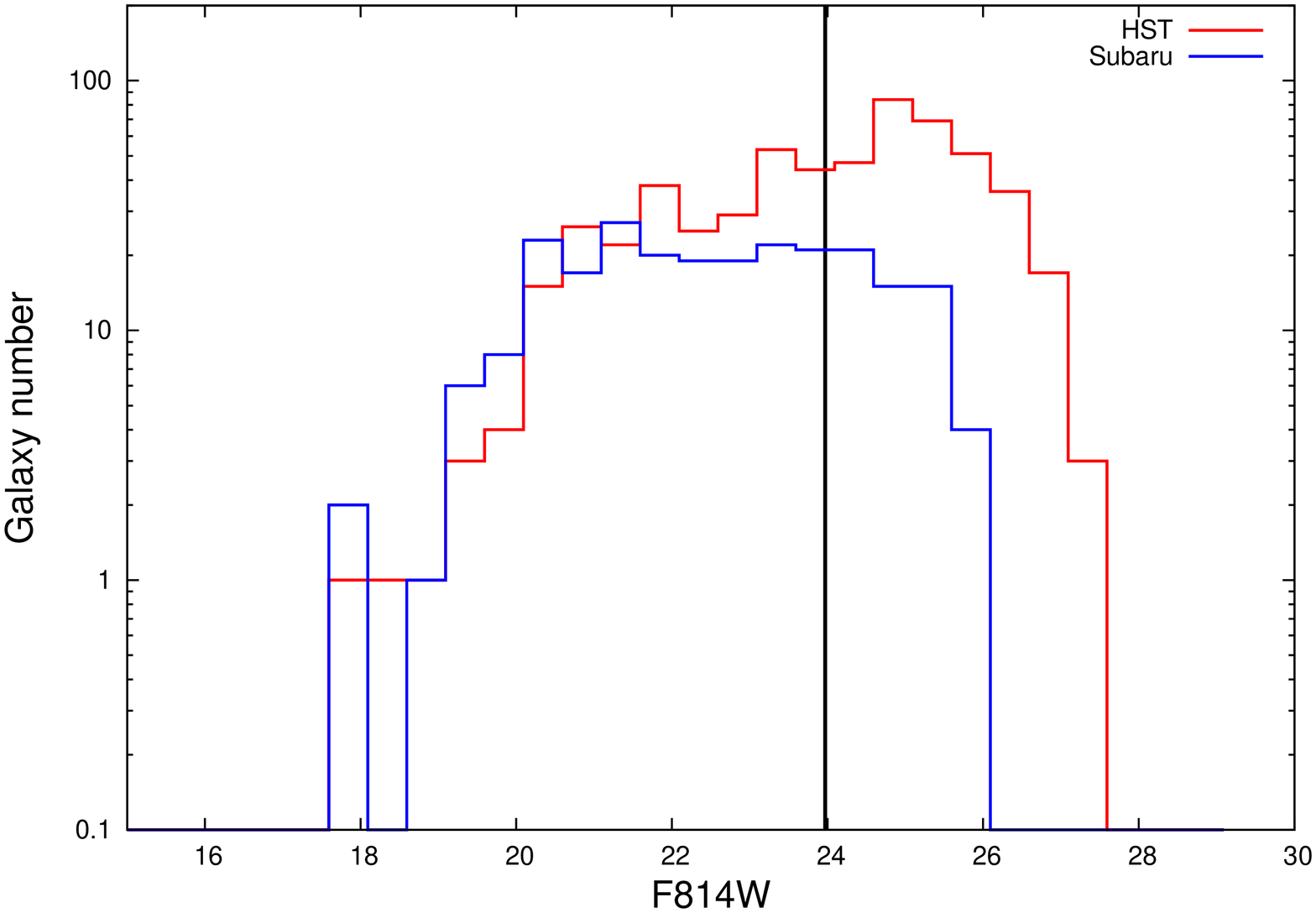}
\caption{Color-magnitude diagrams for RXJ1347, based on the HST
  F606W/F814W data ({\it top}) and on the Subaru V/Ic data ({\it
    middle}). Green crosses correspond to the galaxies before the
  selection, and red circles to the selected RS galaxies. The blue
  lines correspond to the fitted RS and its $\pm 0.3$ color
  width. The {\it bottom} panel shows the distribution of
    magnitudes for the RS galaxies for RXJ1347 with HST data in red
    and Subaru data in blue. The black vertical line indicates the 90\%
    Subaru completeness magnitude limit.}
\label{fig:cmr}
\end{figure}

It has been known for a long time that in a color-magnitude diagram
cluster galaxies follow a relation now called the red sequence, which
has a very small scatter \citep{Bower+92}. The RS
therefore makes the selection of cluster galaxies straightforward when
the filter pair samples the 4000~\AA\ break \citep[see e.g. ][]{Gladders+00}.
We plot a F606W-F814W
versus F814W diagram for HST and V-Ic versus Ic or V-Ip versus Ip for
Subaru. These filters are chosen to bracket the   4000~\AA~break at
the cluster redshifts ($0.187<z<0.686$), highlighting the red
sequence. The RS is fitted by a linear function with a slope fixed
to $-0.0436$ \citep{Durret+11}. We use the same slope for every cluster
as it has been shown that the slope does not significantly vary in this
redshift range \citep[e.g. ][]{DeLucia+07}, which is also what we
observe in our color-magnitude diagrams. The zeropoint is first set to
the early type galaxy color value in \citet{Fukugita+95} at the
cluster redshift, and then re-evaluated in a fit to the RS with a
width of $\pm0.6$ in color and using only galaxies brighter than
I=23. The final width of the RS is set to $\pm0.3$, a classical value in the literature \citep[e.g. ][]{DeLucia+07, Martinet+15a}. Additionally, \citet{Durret+16} studied the impact of the width of the RS on cluster member selection, and found that this value is a good trade off between including many cluster galaxies and limiting the  contamination from field galaxies. As an example we
show the HST and Subaru color-magnitude diagrams of RXJ1347 in
Fig.~\ref{fig:cmr}. The RS is well defined in both data sets. We
observe a higher number of galaxies in the HST catalog, especially at
the faint end, highlighting the higher sensitivity of space-based telescopes, which can be in part attributed to the SB selection effect
described and simulated in \citet{DePropris+13}, and shown in Fig.~\ref{fig:redetect} of the present analysis. We note however that
the field images are affected in the same way, so that more faint
field galaxies are observed in HST than in Subaru.

Selecting cluster galaxies through the RS, while requiring only two
optical bands, does not allow us to study the blue, often late type, cluster
galaxies, which lie below the RS. This can be achieved when galaxy
redshifts are available, and has been applied using photometric
redshifts in e.g. \citet{Martinet+15a}. In the present study we will
therefore only compute the GLFs for the RS cluster galaxies.

\subsection{Computing rest-frame magnitudes}
\label{subsec:rest}

We use LePhare \citep{Arnouts+99,Ilbert+06} to compute k-corrections
and magnitude shifts to obtain rest-frame magnitudes in the F814W and
F606W optical bands for both space and ground based data. Galaxy spectral
energy distributions (SEDs) are modeled with emission lines from
\citet{Polletta+06,Polletta+07} and extinction laws from
\citet{Calzetti+99}. We then select early type galaxy templates at
$\pm0.05$ around each cluster redshift, as those are the most
representative of the RS cluster galaxy population. The k-correction
and magnitude shifts are set to the mean value over the selected
sub-sample of galaxy templates.

We note that RS cluster galaxies have all the same redshifts and
similar colors. Therefore, their k-corrections are similar, and we can apply
the values computed above to every RS galaxy of each cluster. One
consequence is that the histogram of magnitudes for the RS
galaxies is not distorted but simply shifted when going from apparent
to rest-frame magnitudes. This approach assumes that galaxies with the same redshift and V-I color have the same k-correction, which is found to be reasonable in the Sloan Digital Sky Survey (\url{www.sdss.org}) galaxies \citep{Chilingarian+10}. It additionally supposes that the k-corrections are similar across the RS width. Using the online k-correction calculator of \citet{Chilingarian+10, Chilingarian+12} for galaxies below $z=0.5$, we estimate the variation of the k-correction across the $\pm 0.3$ color scatter in the case of RXJ1347. We find a corresponding k-correction scatter of $\pm0.1$ in the I band and $\pm 0.25$ in the V band. We note that these values remain small and do not significantly bias our GLFs which are binned in 0.5 magnitudes. This method is also the best we can do given that we have only two optical bands and therefore cannot perform a proper SED fitting.

\subsection{Subtracting field galaxies}
\label{subsec:field}

Field galaxies are extracted from the COSMOS field the same way as cluster galaxies. Galaxies
are separated from stars in a maximum SB versus magnitude diagram up
to F814W=23 for HST and Ip=22 for Subaru. For each cluster we apply
the same color cut to the field galaxies to select only the galaxies
that lie in the cluster RS. We then apply the same k-correction as
that computed from the cluster galaxy templates. This approach gives
wrong rest-fame magnitudes for the field galaxies because they can be
at different redshifts from that of the cluster. However, as the
k-correction is assumed to be the same for all selected cluster galaxies, removing
 field galaxies k-corrected in this way is equivalent to removing field
galaxies in apparent magnitude. Normalized field galaxy counts are
thus removed from cluster galaxy counts, applying the same
k-correction to both samples and binning in slices of 0.5 magnitude. 

We recall that the COSMOS area used in the present study is $\sim0.05$~deg$^{2}$ which is more than ten times larger than any of the cluster area. We investigate possible cosmic variance effects by computing the galaxy number counts as a function of magnitude in 8 independent subareas of our background catalog. Each of these subregions covers one tenth of the full catalog, and is representative of a typical cluster area. We find no significant deviation of the normalized galaxy counts in these subregions from the counts in the full background catalog. This approach underestimates the cosmic variance, that could still affect our results at larger scales, but we cannot probe it with the small area of our background images. For example \citet{Muzzin+13} showed that even the full COSMOS field has a void at $z\sim1$, and is therefore subject to cosmic variance.

\subsection{Building cluster GLFs}
\label{subsec:build}

Galaxies are counted in bins of 0.5 magnitude and normalized to one
square degree, accounting for masked areas. Error bars are Poisson
errors, and correspond to the quadratic sum of the errors on cluster
and field counts. The normalization is done after computing the errors
to avoid artificially decreasing the error bars.

\subsection{Measuring the completeness}
\label{subsec:comp}

Completeness is a crucial point in GLF studies. Overestimating the
completeness limit can lead to wrong low faint counts, while
underestimating it can mask possible decreases at the faint end. We
note that an overestimation is worst because it introduces bins with
wrong number counts while the underestimation only artificially
degrades the depth of the data. Hence, it is better to adopt a
conservative approach when estimating the completeness limit.

The completeness is measured independently for each image, using
simulated stars. We apply the same code as in \citet{Martinet+15a}. We
first use our measured PSF to model a set of stars with various
magnitudes, and a Gaussian 2D SB profile. We then implement these
stars into the original image and try to re-detect them using
SExtractor with the same configuration file than that used for the
object detection (see Sect.~\ref{subsec:detection}). Doing so with a
thousand stars for each bin of magnitude allows to accurately
determine the completeness of the data. The 90\% completeness limit is
set to the last bin of magnitude at which we still re-detect 90\% of
the simulated stars, minus 0.5 to take into account the fact that
stars are easier to detect than galaxies. \citet{Adami+06}
  estimated the 90\% completeness level for point-like sources and for
  low surface brightness galaxies in their data, and compared these
  limits with a deeper catalog. This led them to ``assume for galaxies
  a mean completeness 0.5 mag brighter than the point source 90\%
  completeness levels, whatever the band'' \citep{Adami+07}. The
  positions at which simulated stars are implemented in the images are
  chosen randomly. Therefore some stars could fall on existing objects
  biasing the estimate of the completeness limit. This bias should
  however be small given the small area of the images covered by
  objects compared to empty regions.

While we measure the completeness for each image, we choose to use the
Subaru completeness limits for the deeper HST images, in order to
compare GLFs from both telescopes in the same magnitude range. We
note that in most cases, this does not significantly affect the fits
to the HST GLFs because the Subaru completeness limit is already
several bins deeper than the characteristic magnitude $M^*$, though
the errors on the parameters are slightly degraded due to the loss
of the faintest bins.

\subsection{Fitting cluster GLFs}
\label{subsec:fit}

Cluster GLFs are fitted with a Schechter function \citep[][;
  eq.~\ref{eq:schech}]{Schechter76}:

\begin{equation}
\label{eq:schech}
 N(M)  = 0.4\ln(10)\phi^*[10^{0.4(M^*-M)}]^{(\alpha+1)}\exp(-10^{0.4(M^*-M)}),
\end{equation}

\noindent where $M^*$ is the characteristic absolute magnitude at
which the GLF bends from bright to faint galaxies, $\alpha$ the
faint-end slope of the GLF, and $\phi^*$ a normalization factor.

We evaluate these three parameters by minimizing the $\chi^2$ between
the Schechter function and the data up to the completeness
limit. Parameter error bars correspond to the 1$\sigma$ confidence level,
and are computed from the covariance matrix, evaluated at the best
parameter values. The final $\chi^2$ value of the fit is converted
into a confidence probability $p$ assuming a $\chi^2$ distribution with three degrees of freedom ($\alpha$,$M^*$,$\phi^*$). This probability of the $\chi^2$ to be lower than the measured value is equal to the incomplete gamma function estimated at $(\chi^2/2,\nu/2)$, where $\nu$ is the number of degrees of freedom. For $\nu=3$, we find:

\begin{equation}
p(\chi^2,\nu) = \frac{2}{\sqrt{\pi}} \left[ \frac{\sqrt{\pi}}{2} {\rm erf}\left(\sqrt{\frac{\chi^2}{2}}\right) - \exp{\left( -\frac{\chi^2}{2} \right)} \sqrt{\frac{\chi^2}{2}} \right].
\end{equation}

There is a known excess of bright galaxies compared to the Schechter
function in the case of clusters. While some authors account for this
excess, for example by fitting a combination of a Schechter and a
Gaussian \citep[e.g. ][]{Biviano+95}, we choose to use a simple
Schechter function for several reasons. First, using a more complex
function with a higher number of parameters decreases the significance
of the fit. This is mainly a concern for the high redshift clusters
which cover fewer bins of magnitude. Second, the very bright end of
the GLF has very high Poisson errors (as high as the signal for the
bin containing the Brightest Cluster Galaxy (BCG)), and thus does not
significantly affect Schechter parameter estimates. We verify this
  statement on the stack GLF estimating Schechter parameters taking or
  not the brightest bins into account, and find a variation of the
  order of 0.1$\sigma$. Therefore, we can safely neglect this excess
when studying the faint end of cluster
GLFs.

\subsection{Stacking cluster GLFs}
\label{subsec:stack}

Individual cluster GLFs are stacked using Colless stacks
\citep[e.g. ][]{Colless89,Martinet+15a}. The idea of this
method is, for each bin of magnitude, to average cluster counts from
every cluster that is 90\% complete at least up to that
bin. Individual counts are first normalized by the cluster richness to
avoid being dominated by large clusters. In this study we define the
richness as the number of galaxies brighter than the brightest
completeness limit of the clusters included in the
stack. To get physical galaxy counts, each final bin value
is multiplied by the mean richness of all clusters included in that
bin.

While this method allows to use the maximum amount of information from
the cluster data set, its interpretation
 requires some care. The main issue is that the number of clusters is
different in each magnitude bin. In particular the faintest bins
are only populated by clusters with the faintest completeness
limit. To avoid the GLF faint end being dominated by one or two very
complete clusters, we only take into account the bins that
include at least four clusters. This number is chosen based on
\citet{Martinet+15a} who found that for a given completeness limit,
the estimated Schechter parameters tend to remain the same when adding
more clusters in the stack. In addition, each bin of magnitude
corresponds to a different mean redshift, which can have important
consequences given the large redshift range of the studied cluster
sample ($0.187\leq z\leq0.686$). Since the absolute magnitude
completeness limit is brighter at higher redshifts (assuming that all
data have approximately the same depth), the faint bins of the GLF
are dominated by the lower redshift clusters. This problem can be
attenuated by stacking clusters in redshift bins, though this decreases the
significance of the signal.

\section{Galaxy luminosity functions: results}
\label{sec:res}

We first show the individual (Sect.~\ref{subsec:indiv}) and stacked
cluster GLFs (Sect.~\ref{subsec:stackk}). We then study the evolution
of the stacked GLFs with redshift (Sect.~\ref{subsec:stackz}) and mass
(Sect.~\ref{subsec:stackM}) for the HST and Subaru data. We also
investigate binning both in redshift and mass to break the degeneracy
between the two parameters (Sect.~\ref{subsec:stackzM}). In the case
of Subaru, we  compute the GLFs as a function of radius as well
(Sect.~\ref{subsec:rad}). Finally, we use simulations to study the effect of SB dimming (Sect.~\ref{subsec:SBdim}). As we found that the studied GLFs behave
identically in the F606W and F814W filters, we only show them in the
latter filter, to improve the paper readability.

\subsection{Individual cluster GLFs}
\label{subsec:indiv}

\begin{figure*}
 \begin{tabular}{cccc}
\includegraphics[width=0.17\textwidth,clip,angle=270]{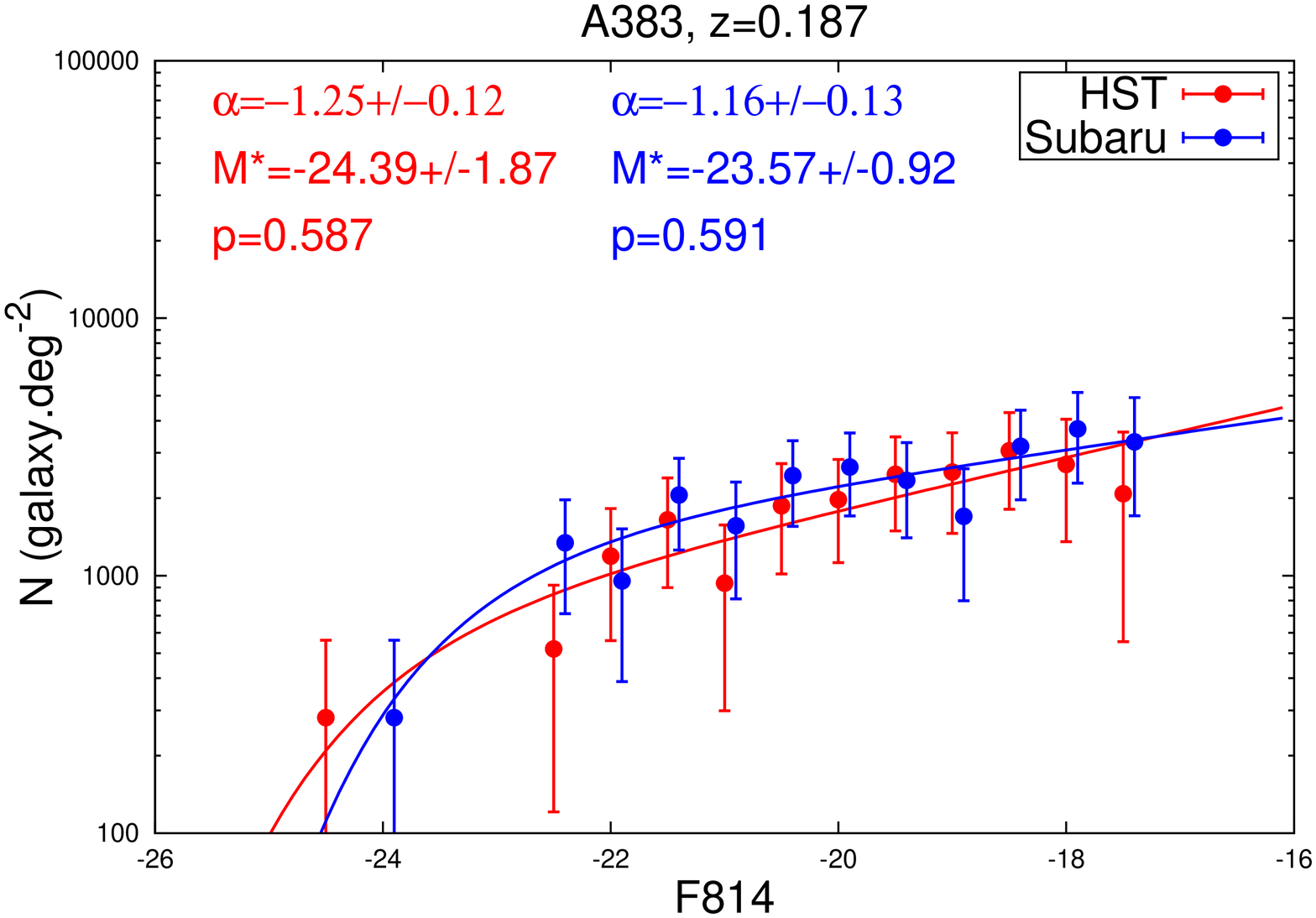}
\includegraphics[width=0.17\textwidth,clip,angle=270]{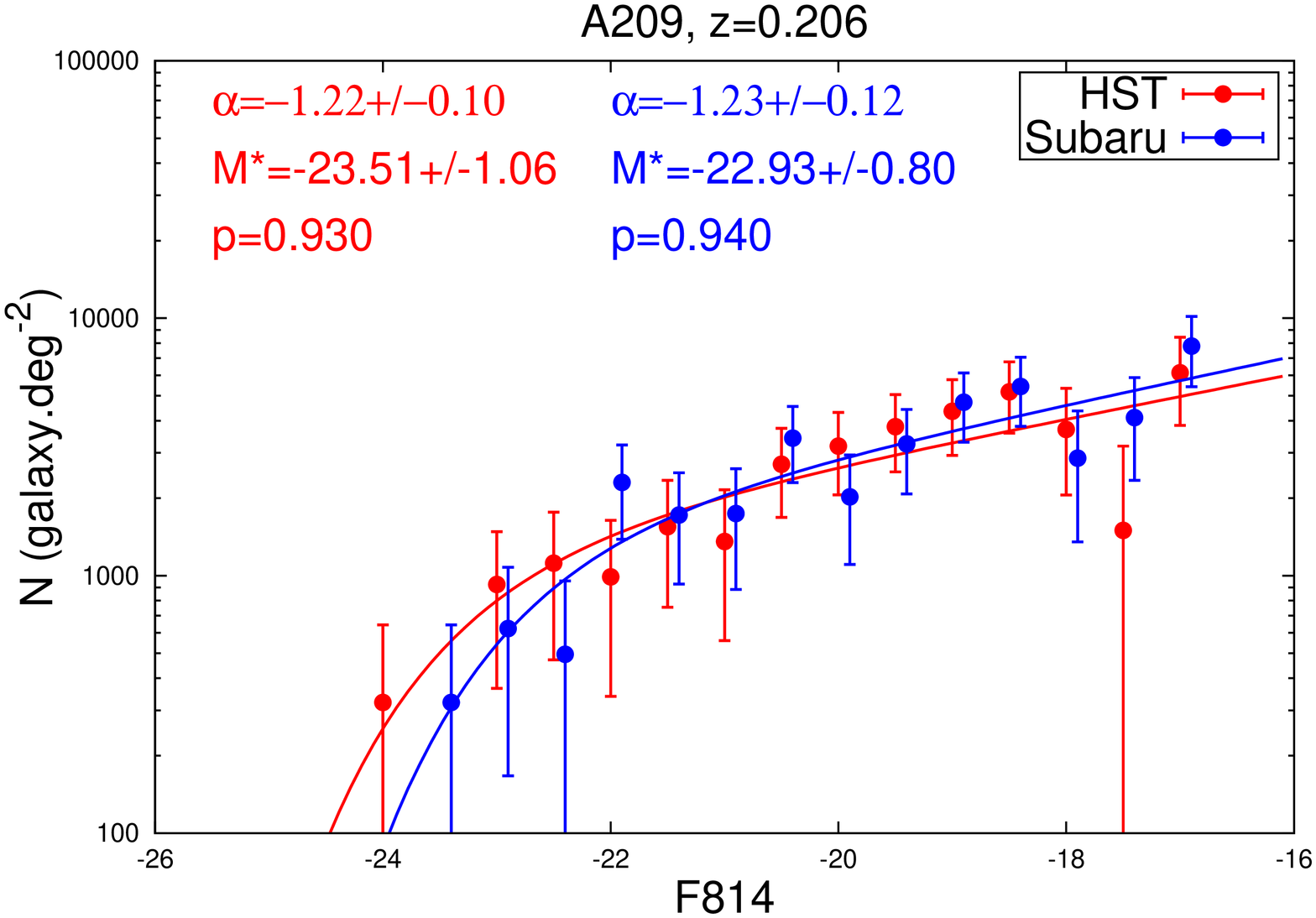}
\includegraphics[width=0.17\textwidth,clip,angle=270]{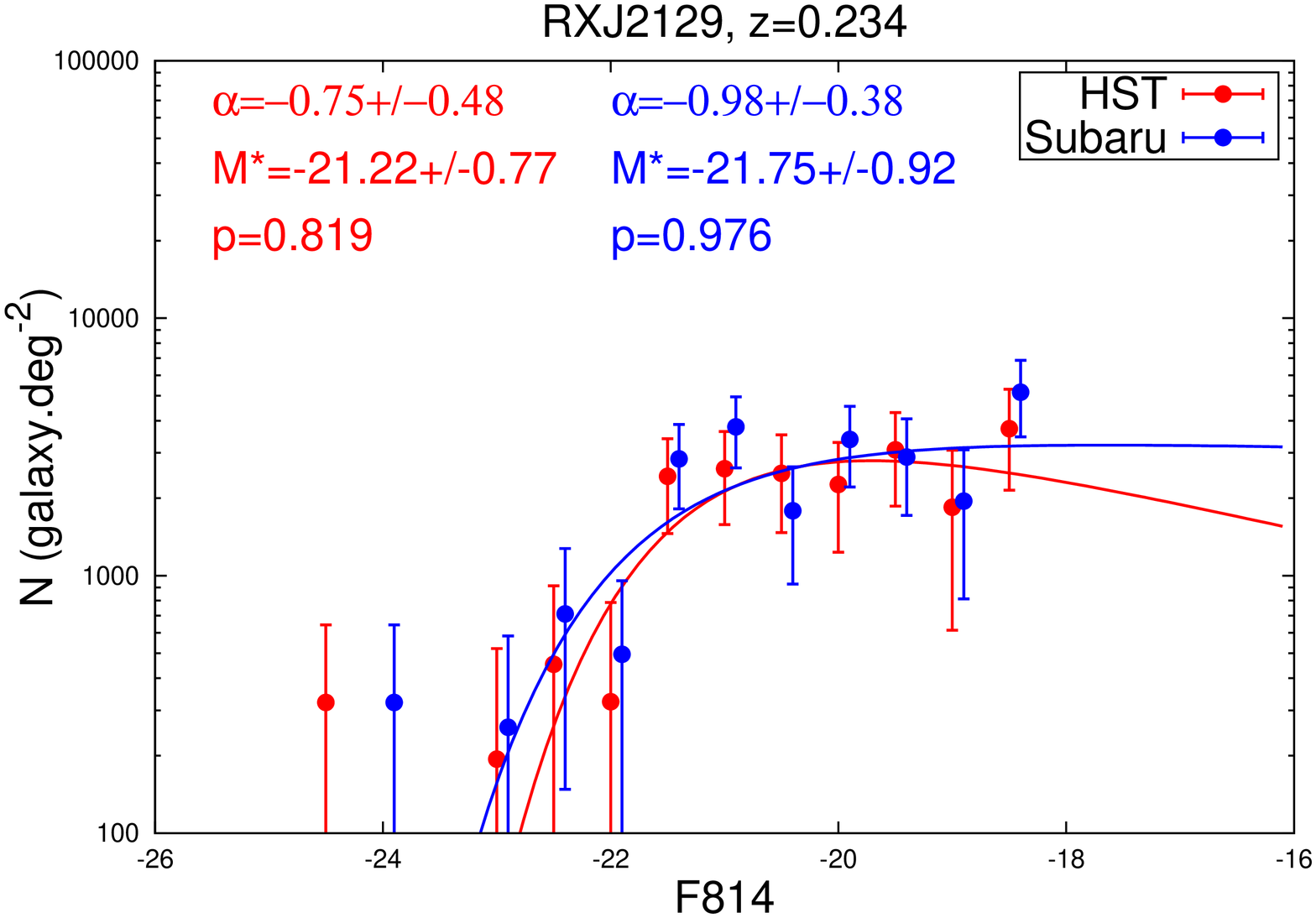}
\includegraphics[width=0.17\textwidth,clip,angle=270]{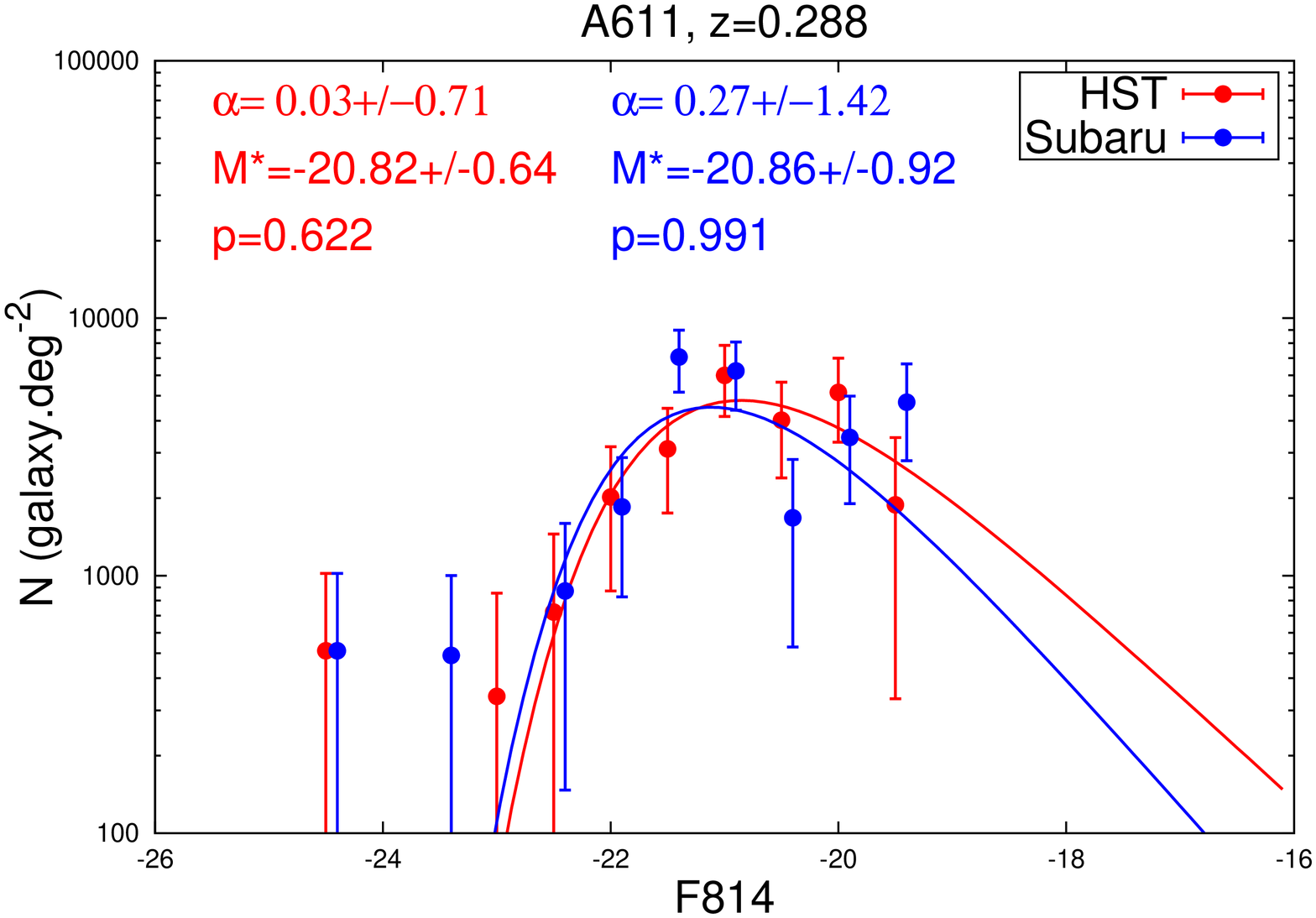} \\
\includegraphics[width=0.17\textwidth,clip,angle=270]{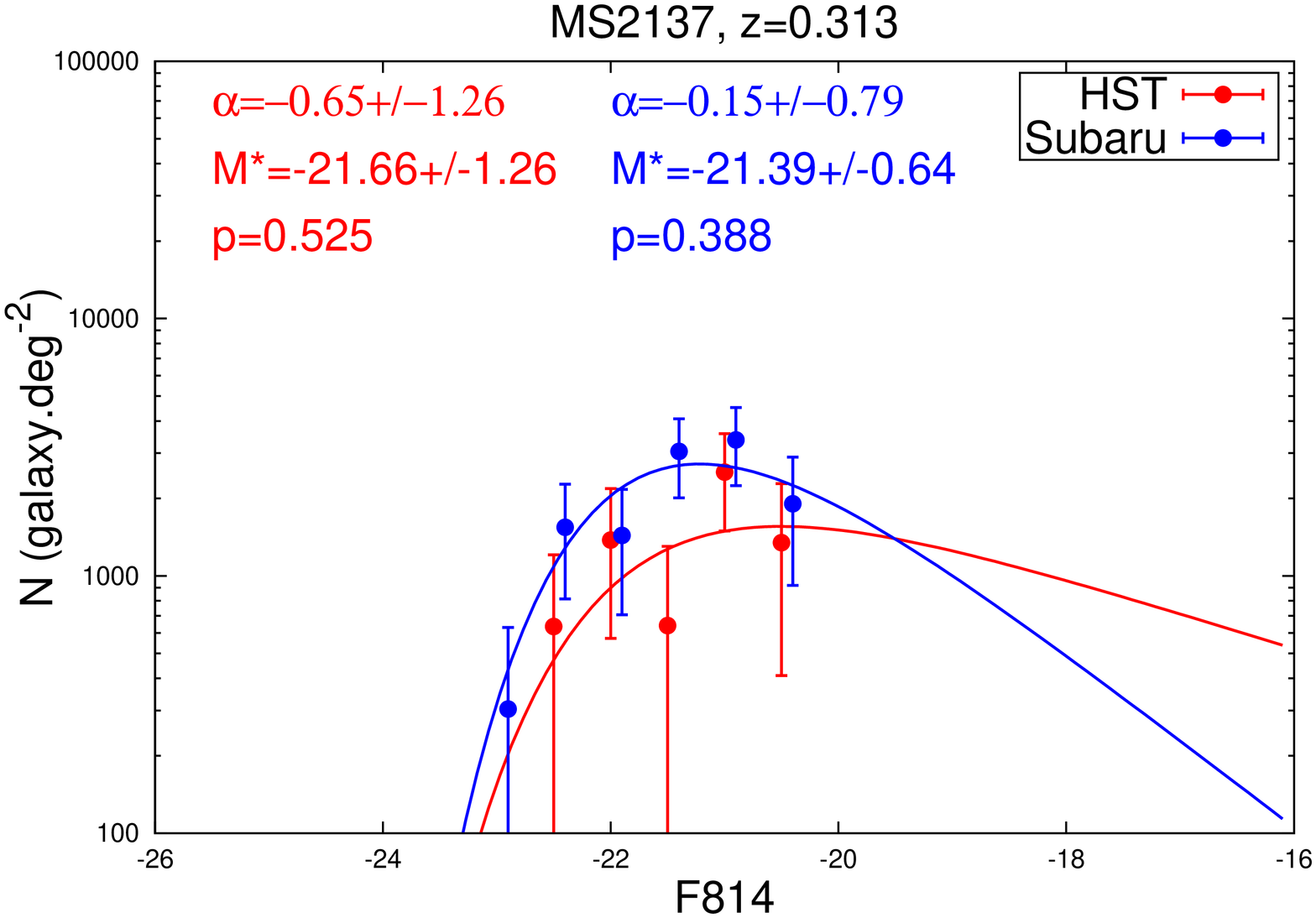}
\includegraphics[width=0.17\textwidth,clip,angle=270]{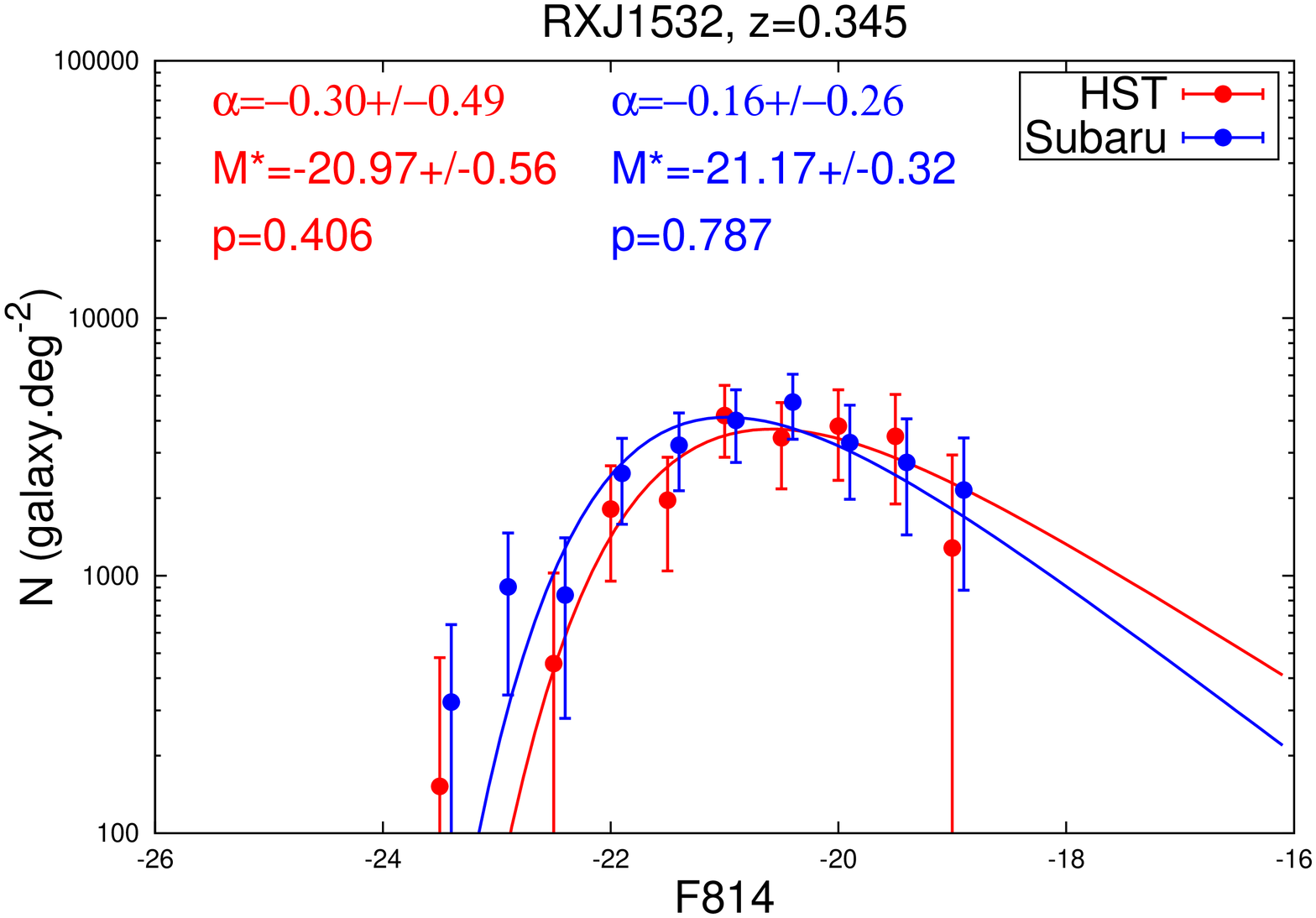}
\includegraphics[width=0.17\textwidth,clip,angle=270]{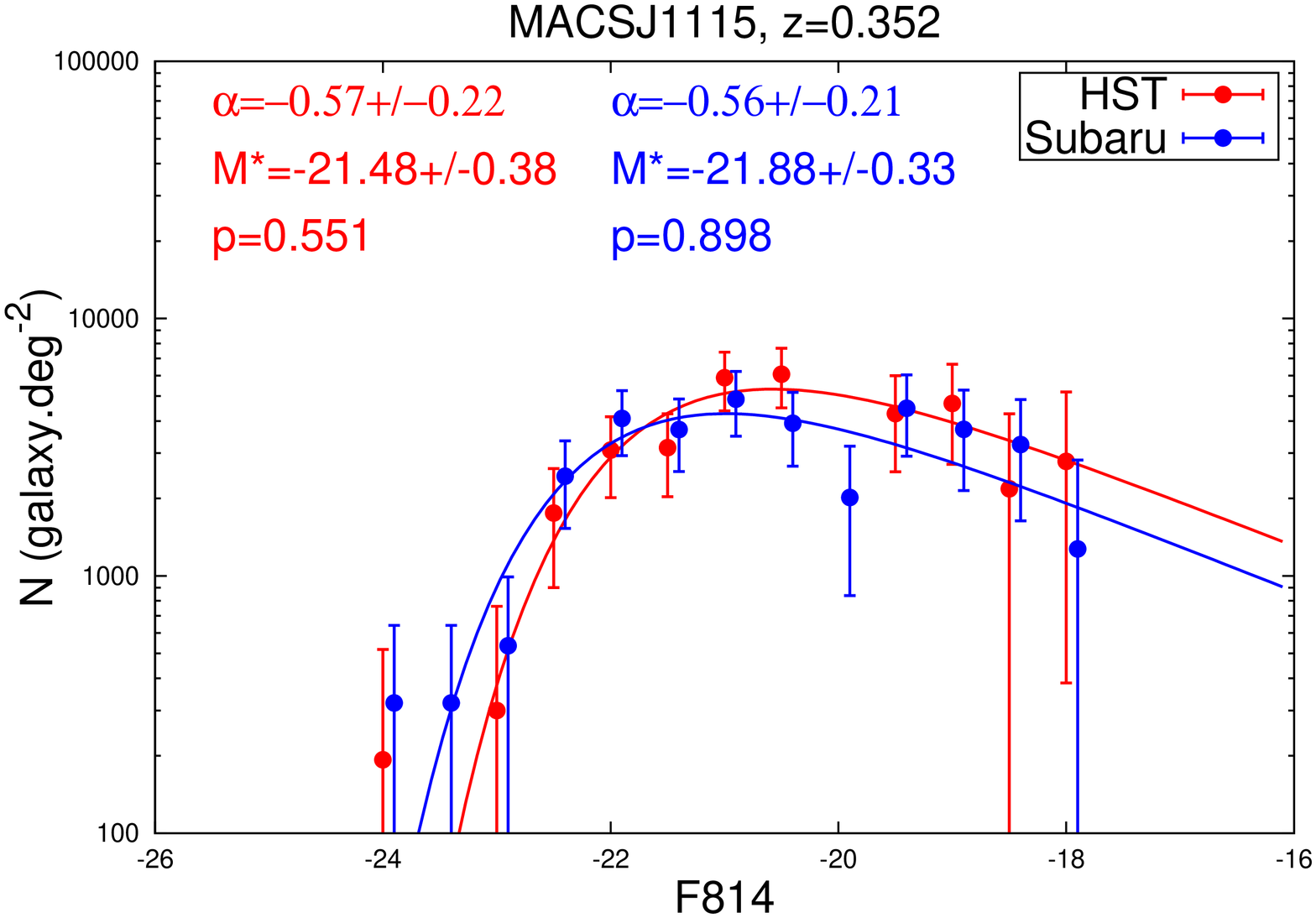}
\includegraphics[width=0.17\textwidth,clip,angle=270]{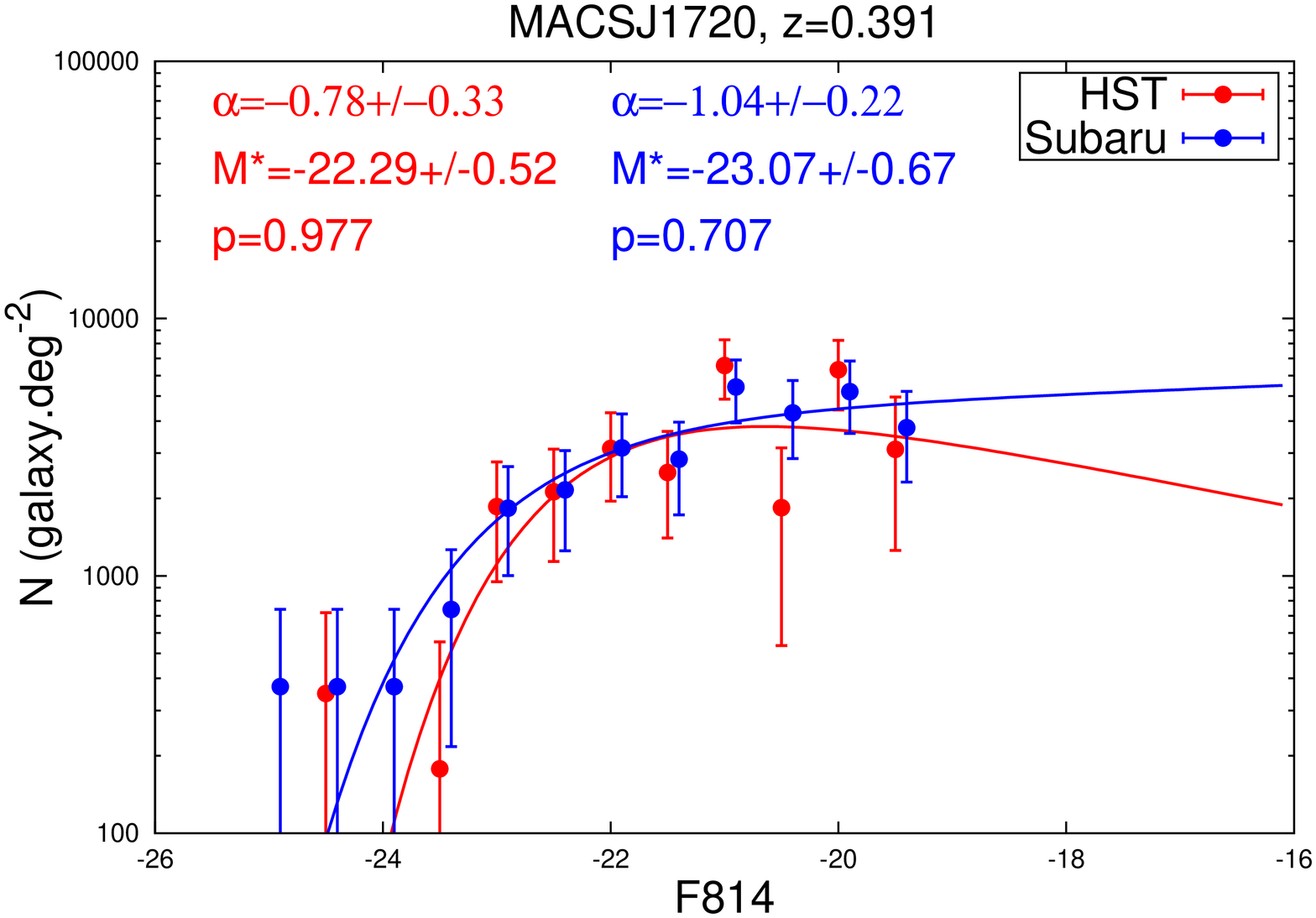} \\
\includegraphics[width=0.17\textwidth,clip,angle=270]{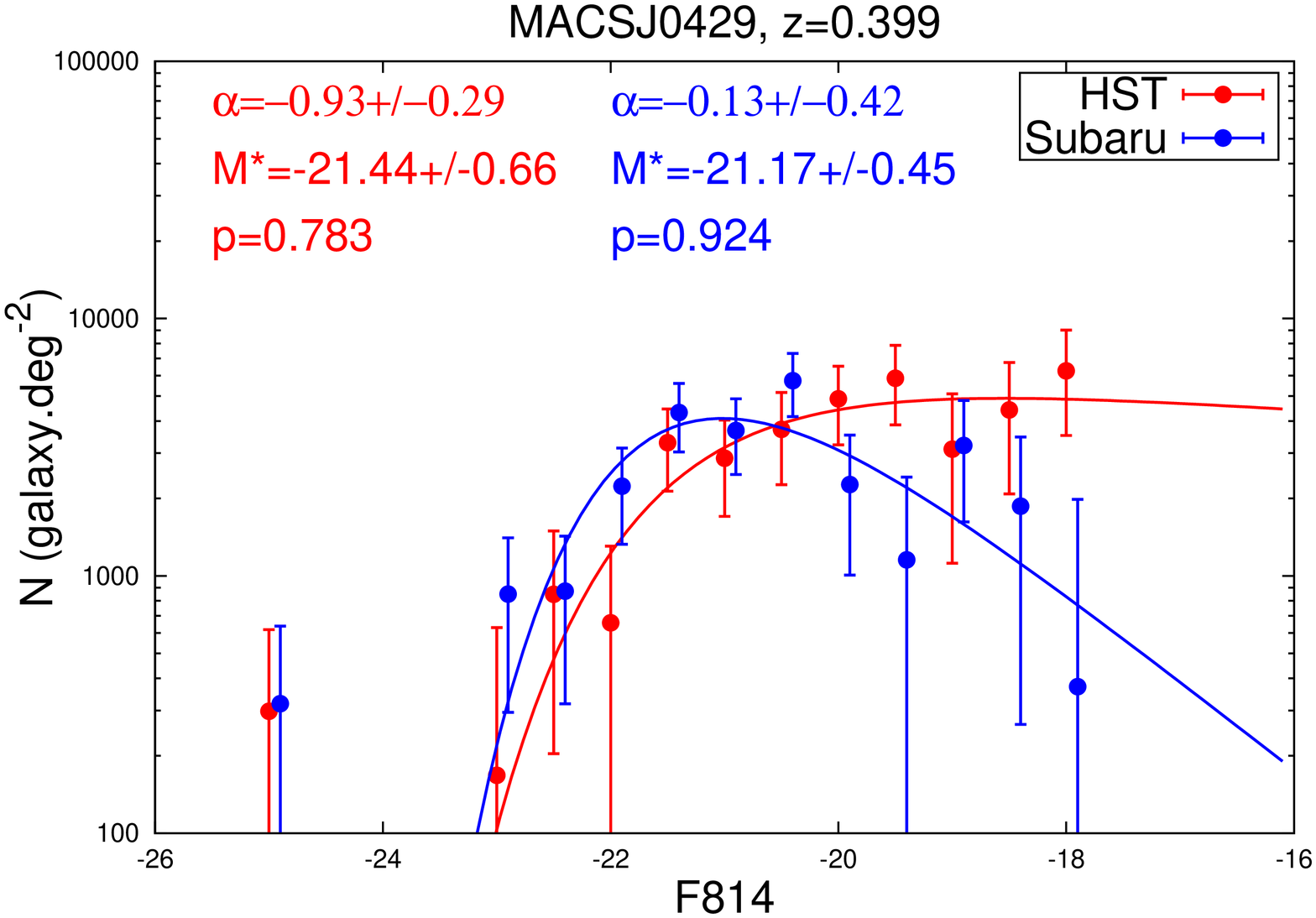}
\includegraphics[width=0.17\textwidth,clip,angle=270]{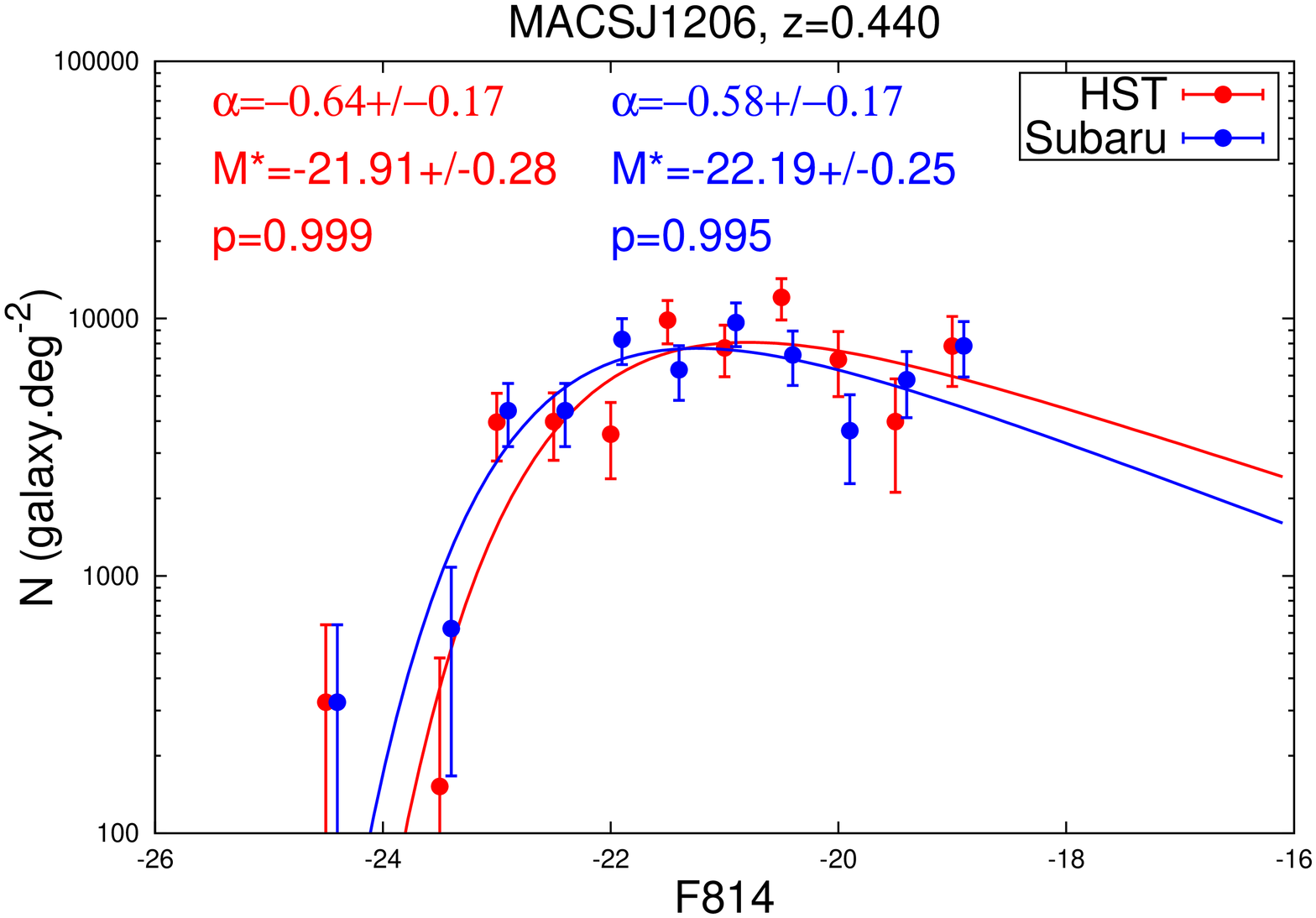}
\includegraphics[width=0.17\textwidth,clip,angle=270]{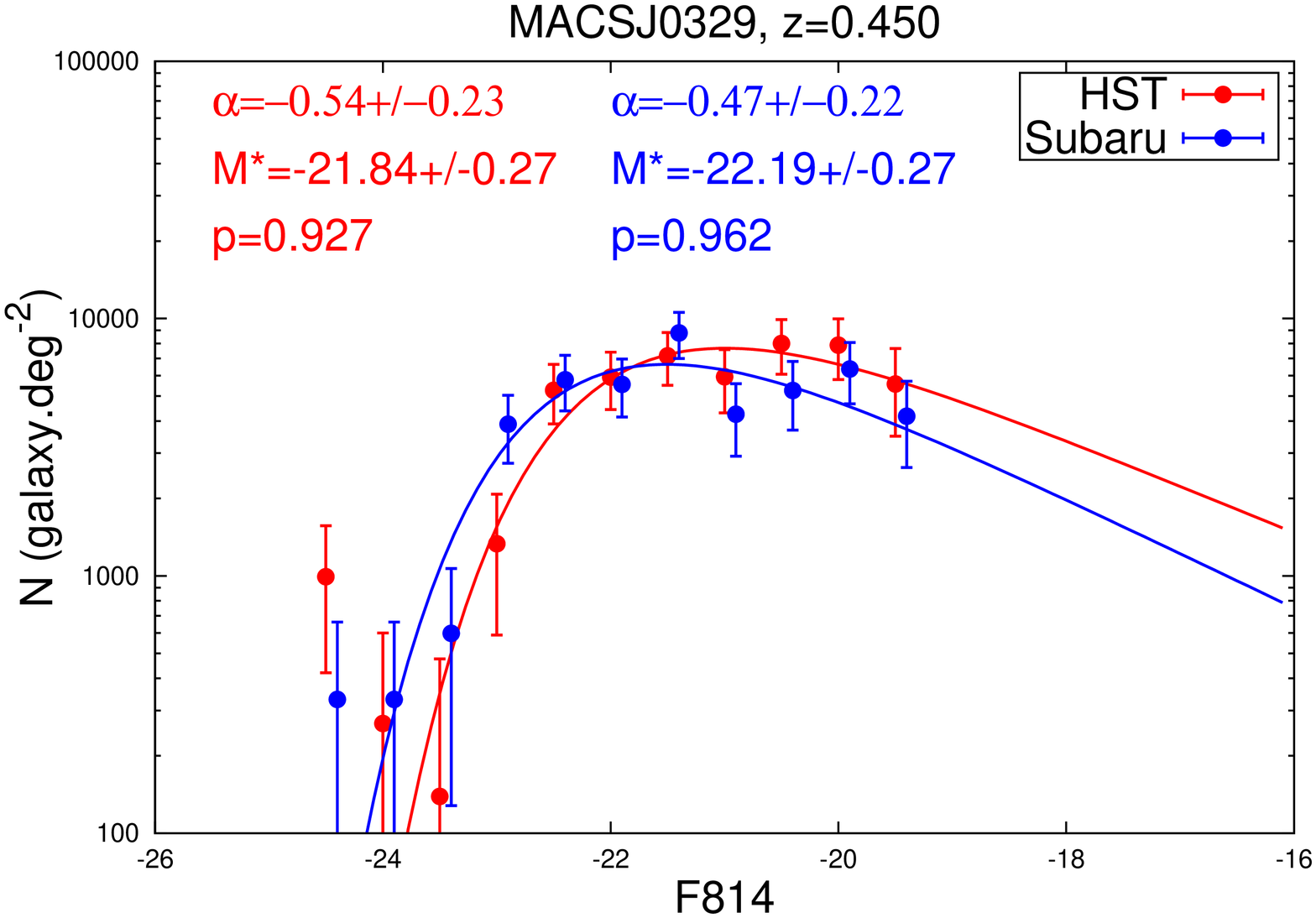}
\includegraphics[width=0.17\textwidth,clip,angle=270]{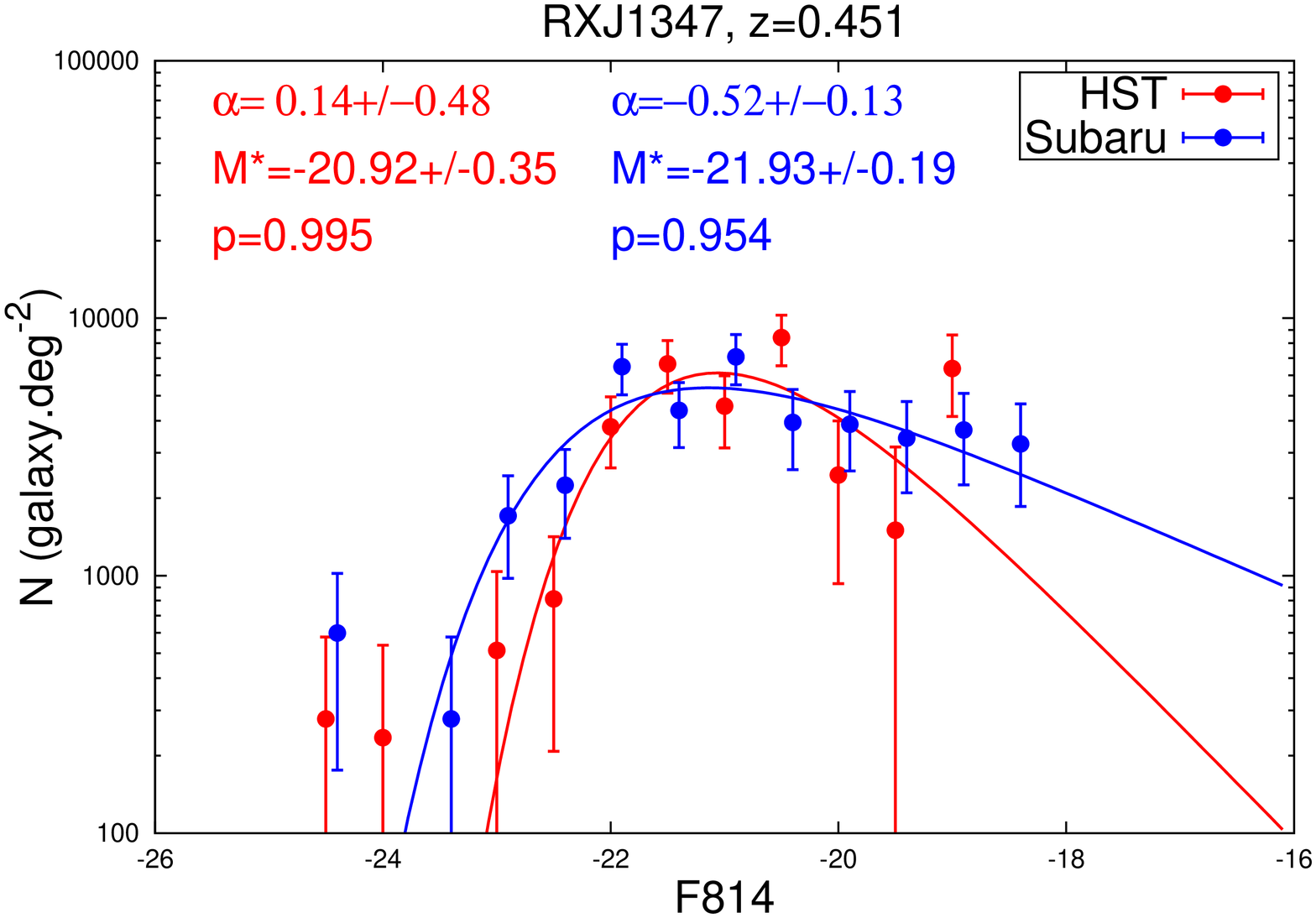} \\
\includegraphics[width=0.17\textwidth,clip,angle=270]{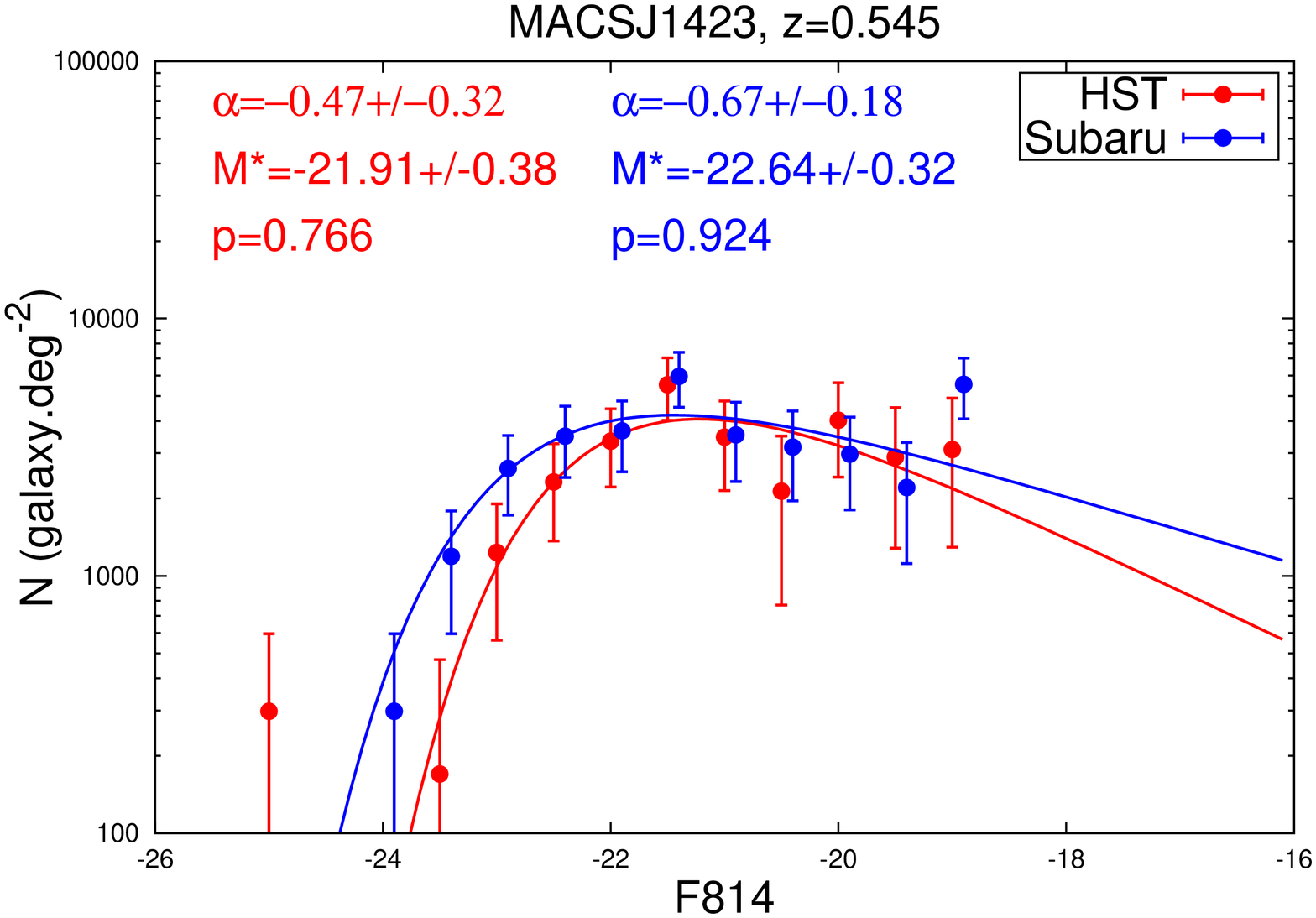}
\includegraphics[width=0.17\textwidth,clip,angle=270]{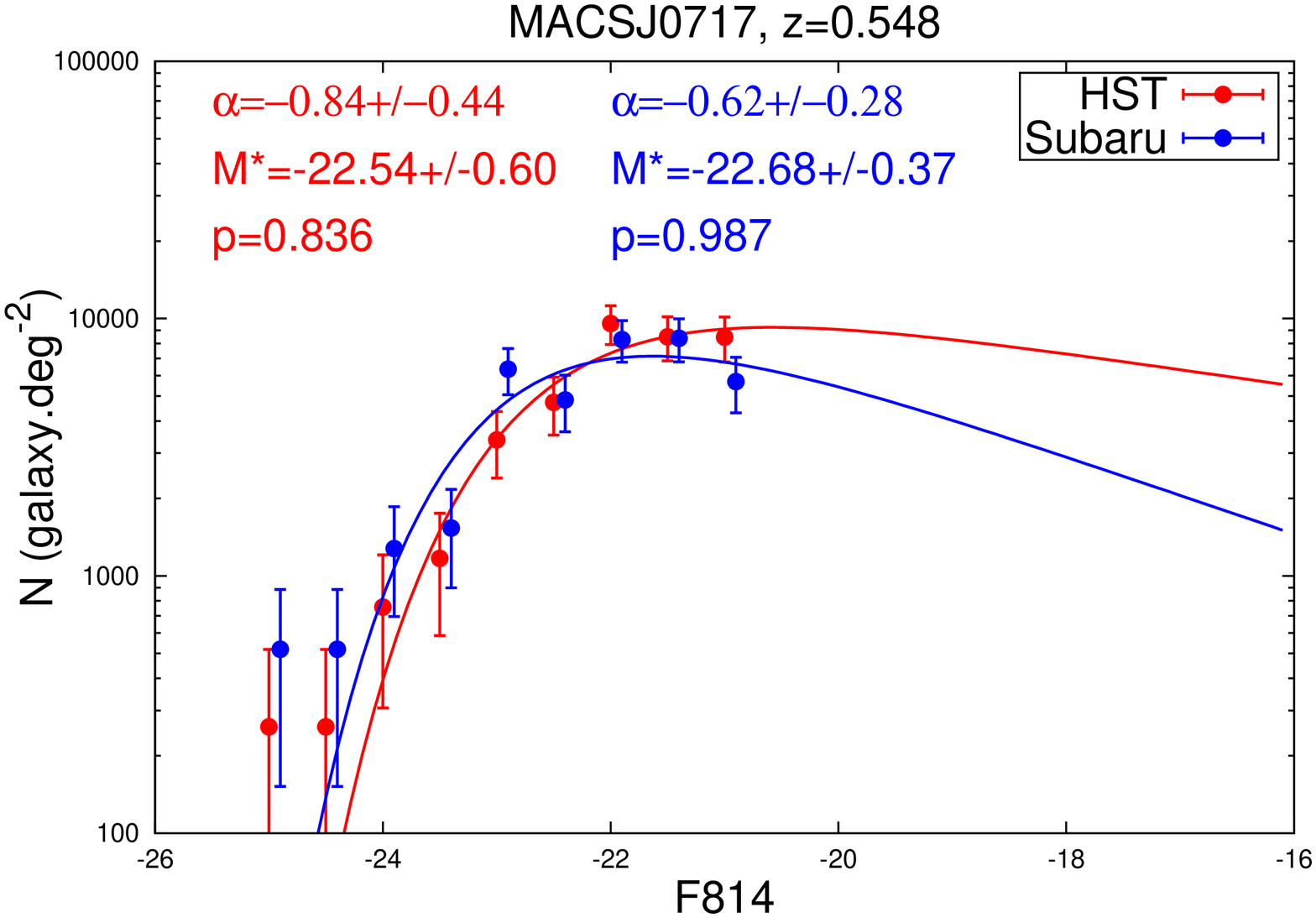}
\includegraphics[width=0.17\textwidth,clip,angle=270]{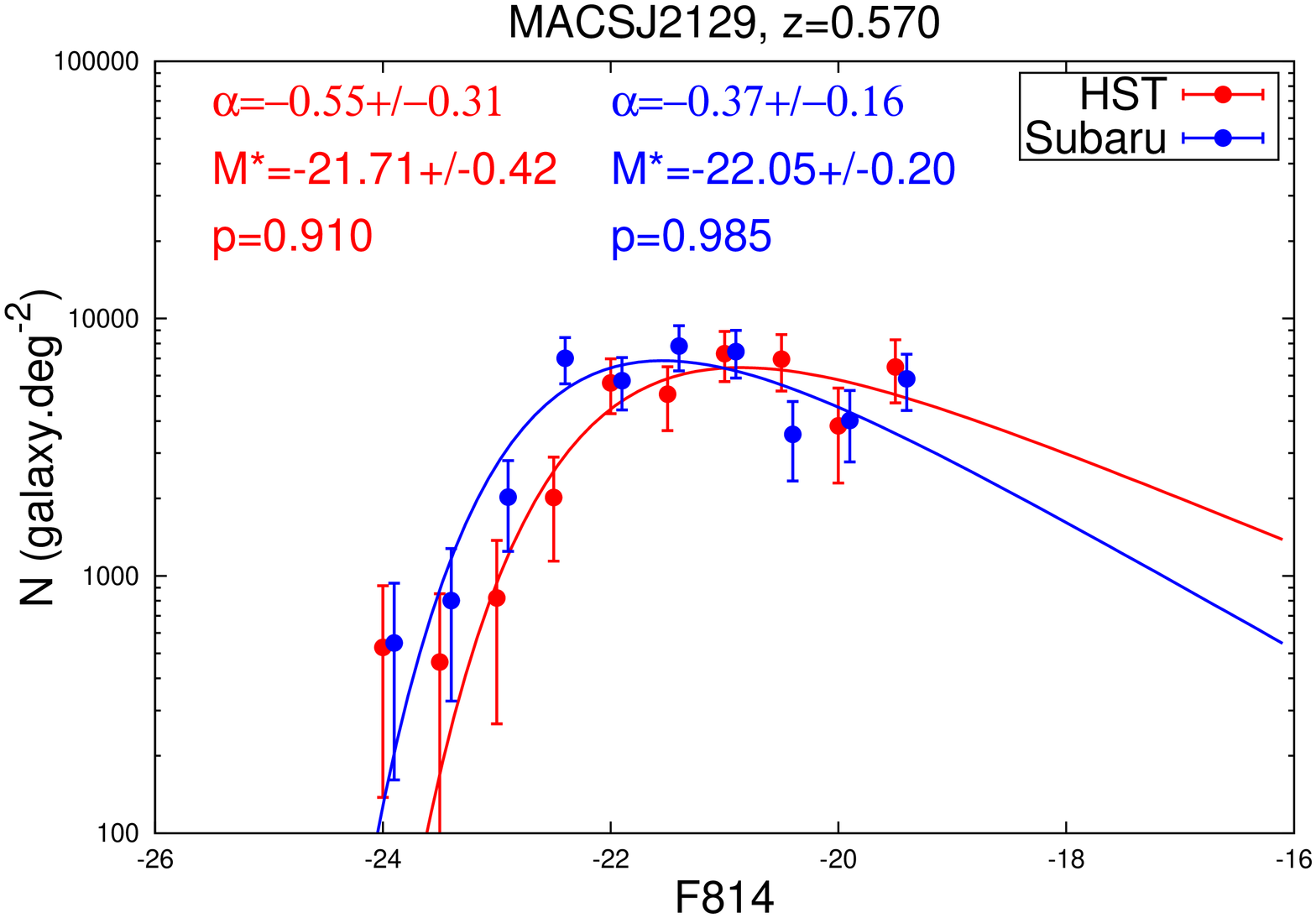}
\includegraphics[width=0.17\textwidth,clip,angle=270]{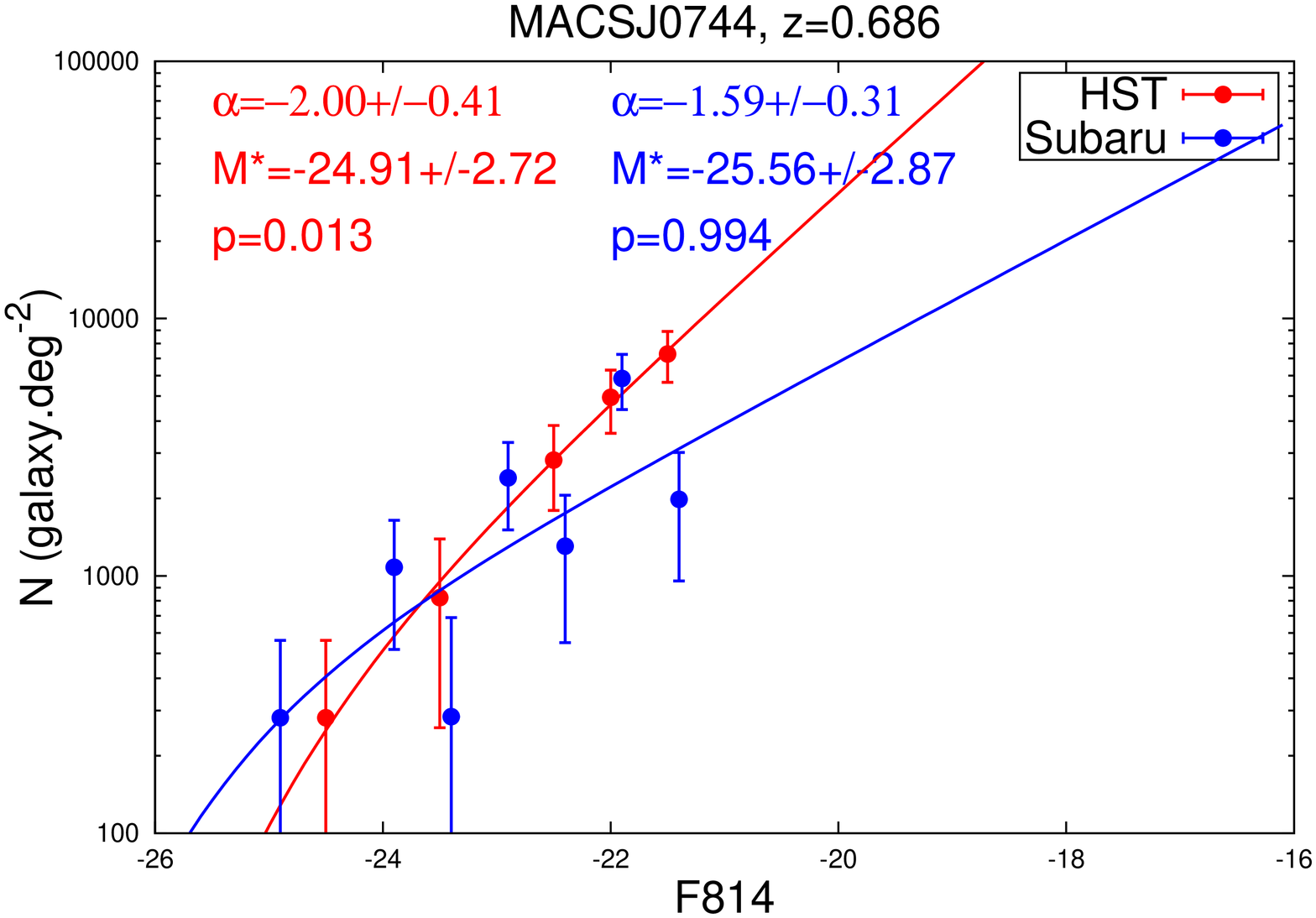} \\
 \end{tabular}
 \caption{Individual cluster GLFs in the F814W filter, sorted from low
   to high redshift. Red and blue correspond to the GLFs measured with
   HST and Subaru respectively. The curves are the Schechter fits
   to the data up to the 90\% completeness magnitude limit. The
   parameters from each fit are displayed in the corresponding
   color: the slope $\alpha$, the characteristic magnitude $M^*$, and the significance of the fit $p$ defined in eq.~(2). Each GLF is computed within a circle centered on the cluster
   center with the largest possible radius given the HST field of
   view, and is normalized to one square degree.}
\label{fig:glfindivf814}
 \end{figure*}

Figure~\ref{fig:glfindivf814} shows the individual cluster GLFs in the
F814W band. We note a good overall agreement between the Subaru (in
blue) and HST GLFs (in red). As a general trend we find that the faint
end seems to be flatter for the lower redshifts. However, the
Schechter parameters from the fits have large error bars, and a large
dispersion across clusters. This highlights the need for stacking to
infer precise results on the faint end slope $\alpha$ and the characteristic
magnitude $M^*$. The Schechter parameters from the fits to individual
clusters are displayed in Appendix~\ref{appendix:glfparams}.

\subsection{Stacked cluster GLFs}
\label{subsec:stackk}

\begin{figure}
\centering
\includegraphics[width=0.34\textwidth,clip,angle=270]{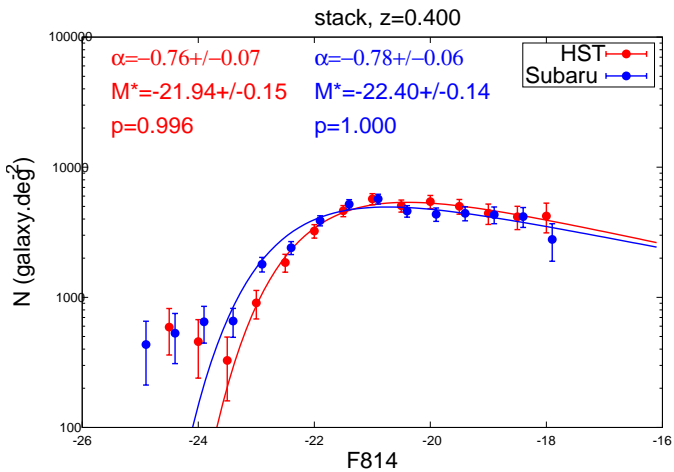}
\caption{Stacked cluster GLFs in the F814W filter. Red and blue
  correspond to the GLFs measured with HST and Subaru respectively,
  and are normalized to one square degree. The curves correspond to
  the Schechter fits to the data. The parameters from each fit are
  displayed in the corresponding color.}
\label{fig:glfstack}
\end{figure}

The stacked GLFs for all 16 clusters are presented in
Fig.~\ref{fig:glfstack}. We find a very good agreement
between the HST and Subaru GLFs, with equal faint end slopes given the error
bars and only a slightly higher $M^*$ for
Subaru. We find $\alpha=-0.76\pm0.07$ for HST and $\alpha=-0.78\pm0.06$ for
Subaru, the mean redshift of clusters being $\bar{z}=0.4$, and
the fit extending to more than $M^*+4$. Even at this depth we cannot investigate the possible
upturn of the GLF which is seen in the very faint population of nearby
clusters \citep[e.g. ][]{Popesso+06}. We also note an expected excess at the very
bright end of the GLF, as discussed in Sect.~\ref{subsec:fit}. GLFs included in
the stack are computed within a circle centered on the cluster center with
the largest possible radius given the HST field of view. Therefore
each cluster covers a different area but we show in
Sect.~\ref{subsec:rad} that the stacked GLFs do not depend on the
radius in which they are computed in the range [0.5:2.5]~Mpc, and with the present completeness limits.

Comparable results are found in the F606W filter, but since the data are not 
as deep as in the F814W filter we do not show them here. 

\subsection{Evolution with redshift}
\label{subsec:stackz}

\begin{figure}
\centering
\includegraphics[width=0.34\textwidth,clip,angle=270]{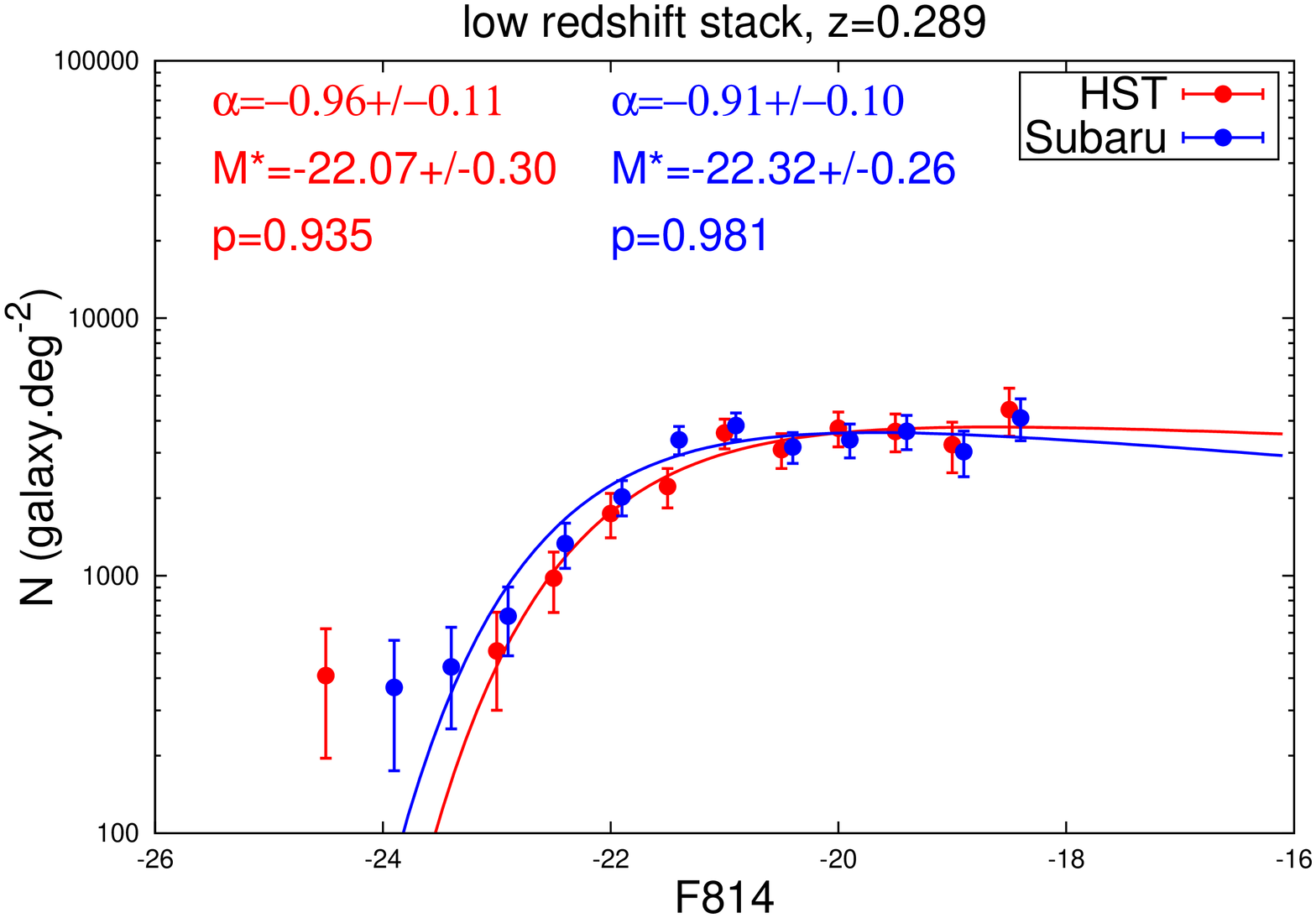}
\includegraphics[width=0.34\textwidth,clip,angle=270]{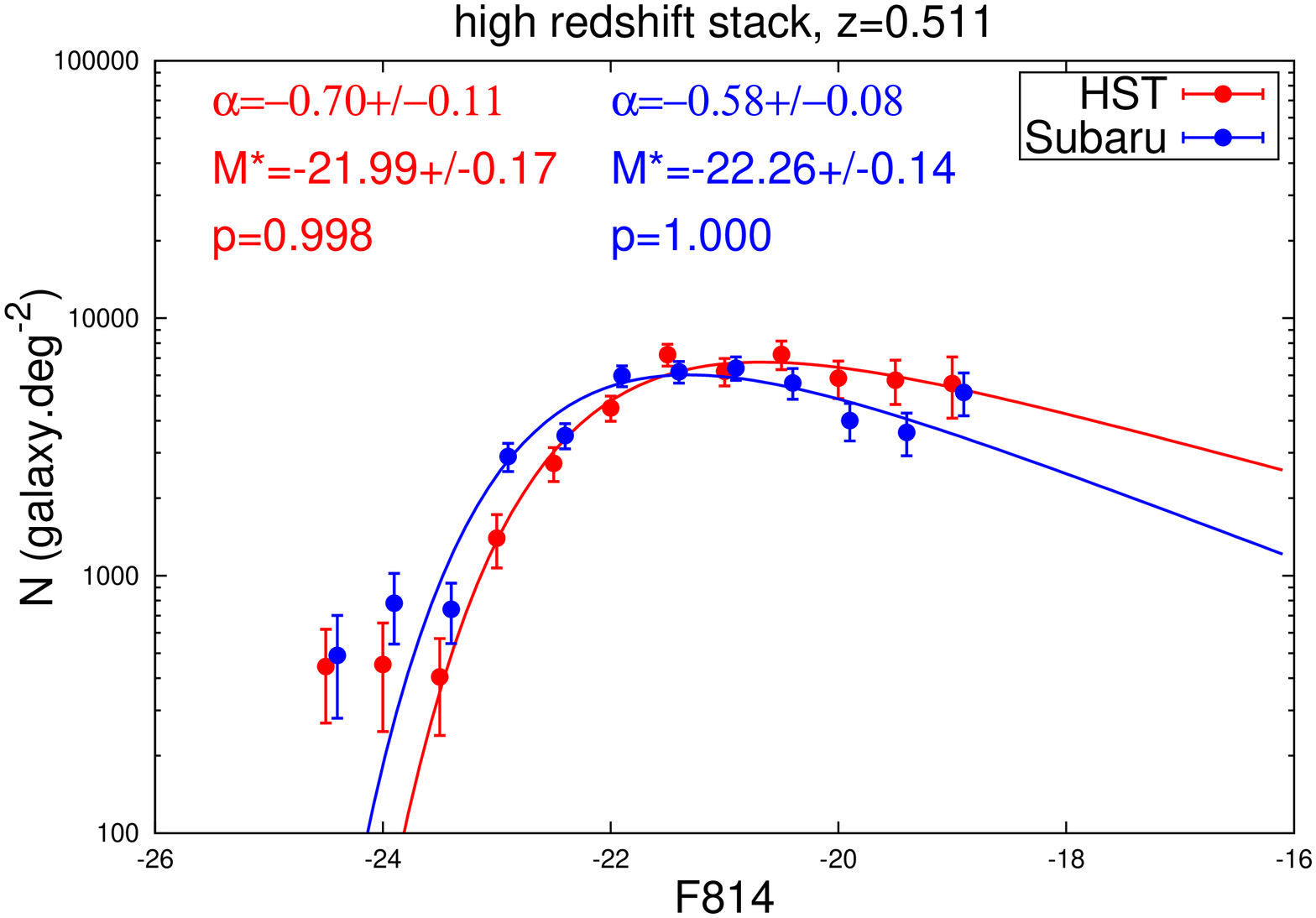}
\caption{Evolution of the stacked cluster GLF with redshift, in the
  F814 filter. {\it Top} represents the low redshift GLF
  ($\bar{z}=0.289$) and {\it bottom} the high redshift
  ($\bar{z}=0.511$). Red and blue correspond to the GLFs
  measured with HST and Subaru respectively, and are normalized to one
  square degree. The curves correspond to the Schechter fits to the
  data. The parameters from each fit are displayed in the
  corresponding color.}
\label{fig:glfstackz}
\end{figure}

As the faint end of the Colless stack is dominated by the most
complete, hence the lowest redshift clusters, we separate our sample
into two  redshift bins. This allows us to better quantify the
evolution of the faint end with redshift. The low redshift sample is
composed of 8 clusters with $0.19<z<0.39$, and the high redshift
sample of 8 clusters with $0.40<z<0.69$. Results are displayed in
Fig.~\ref{fig:glfstackz}.

The redshift segregation highlights some possible differences between
the HST and Subaru GLFs, but we still note a decrease of the faint end
when the redshift increases. In the lower redshift case, both GLFs
agree, and we find $\alpha=-0.96\pm0.11$ and $\alpha=-0.91\pm0.10$ for
HST and Subaru data respectively. We note that the faint end now agrees
with a flat faint end value ($\alpha=-1$), for a mean redshift of
$\bar{z}=0.289$. In the high redshift case, the faint ends
computed with HST and Subaru data differ at a 0.9$\sigma$ level, with
$\alpha=-0.70\pm0.11$ and $\alpha=-0.58\pm0.08$, providing  a hint
of a SB selection effect, but not at a significant level. The change in $\alpha$
between the low and high redshift cases is 1.7$\sigma$ and
2.6$\sigma$ for HST and Subaru respectively. While faint object
selections are different for HST and Subaru, leading to higher faint
galaxy counts in the former instrument at high redshift, we still find
a decreasing faint end with increasing redshift. In Sect.~\ref{subsec:SBdim} we use simulations to check whether the observed evolution with redshift can be attributed to SB dimming.

\subsection{Dependence on mass}
\label{subsec:stackM}

Using total masses computed from joint weak and strong lensing by
\citet{Umetsu+15}, and given in Table~\ref{tab:data}, we can separate
clusters into low mass
($6\times10^{14}M_\odot<M_{200}<10^{15}M_\odot$) and high mass
($10^{15}M_\odot<M_{200}$) samples. Two clusters in the present study
are not part of the \citet{Umetsu+15} sample. However they can be
safely classified as low mass for MACSJ1423 and high mass for
MACSJ2129, according to the strong lensing analysis of
\citet{Zitrin+11}. In addition, \citet{Martinet+16} derived a weak
lensing mass of $M_{200}=(8.8\pm3.3)\times10^{14}M_\odot$ for
MACSJ1423, in agreement with its strong lensing classification. Given
that we only discriminate clusters according to a mass threshold,
accurate masses are not required, providing that the threshold is excluded by the mass error bars, which is the case for most of our clusters. There are 6 low mass clusters and 10
high mass. We could have chosen a mass threshold such that we have 8
clusters in every stack, but this would result in having clusters of
masses $10^{15}M_\odot<M_{200}$ in the low mass sample, while a
  cluster of $10^{15}M_\odot$ is already a very massive cluster. The low-mass sample has a median mass of $M_{200}=7.98 \times 10^{14} M_\odot$, and the high-mass of $M_{200}=16.66 \times 10^{14} M_\odot$.

\begin{figure}
\centering
\includegraphics[width=0.34\textwidth,clip,angle=270]{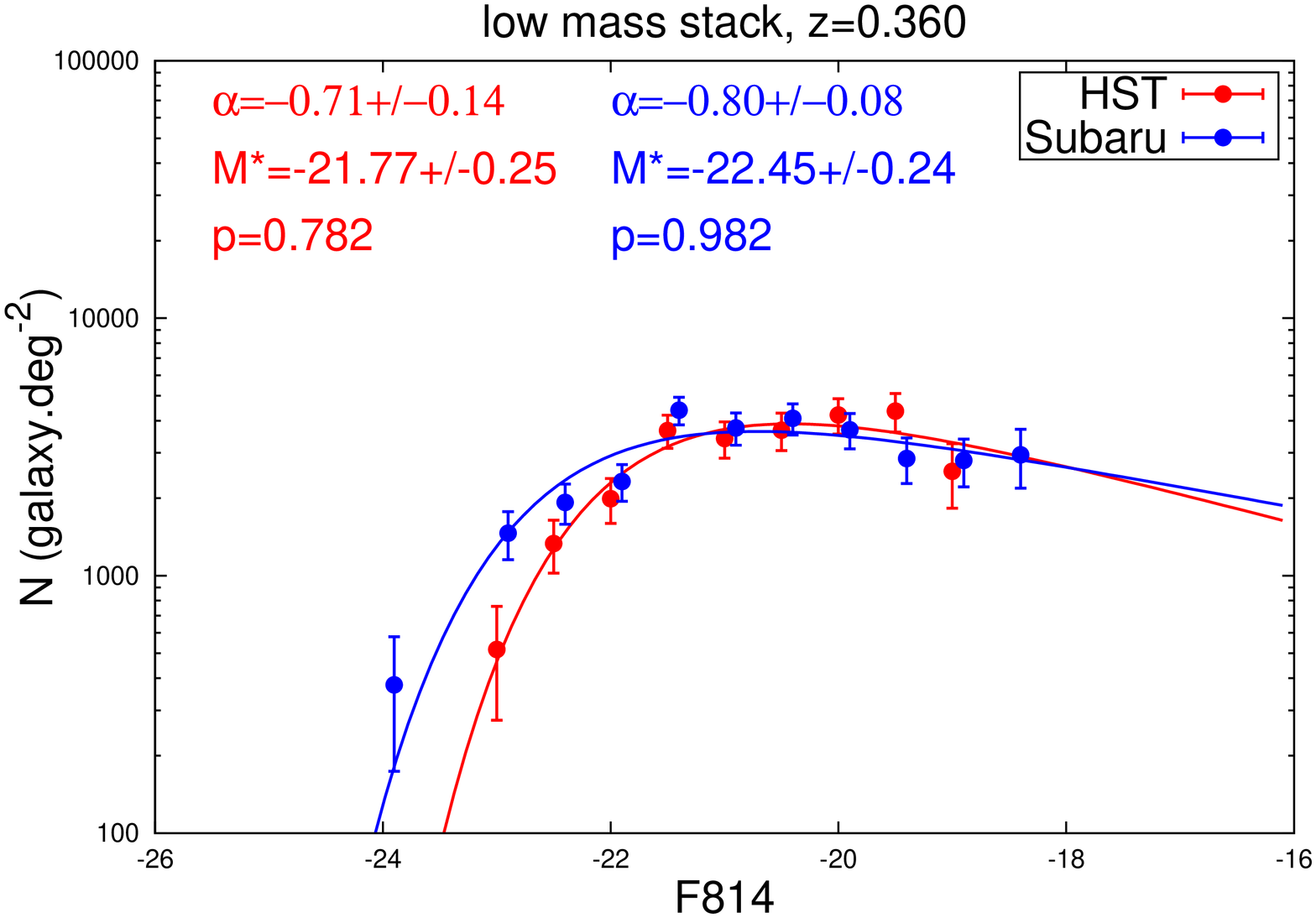}
\includegraphics[width=0.34\textwidth,clip,angle=270]{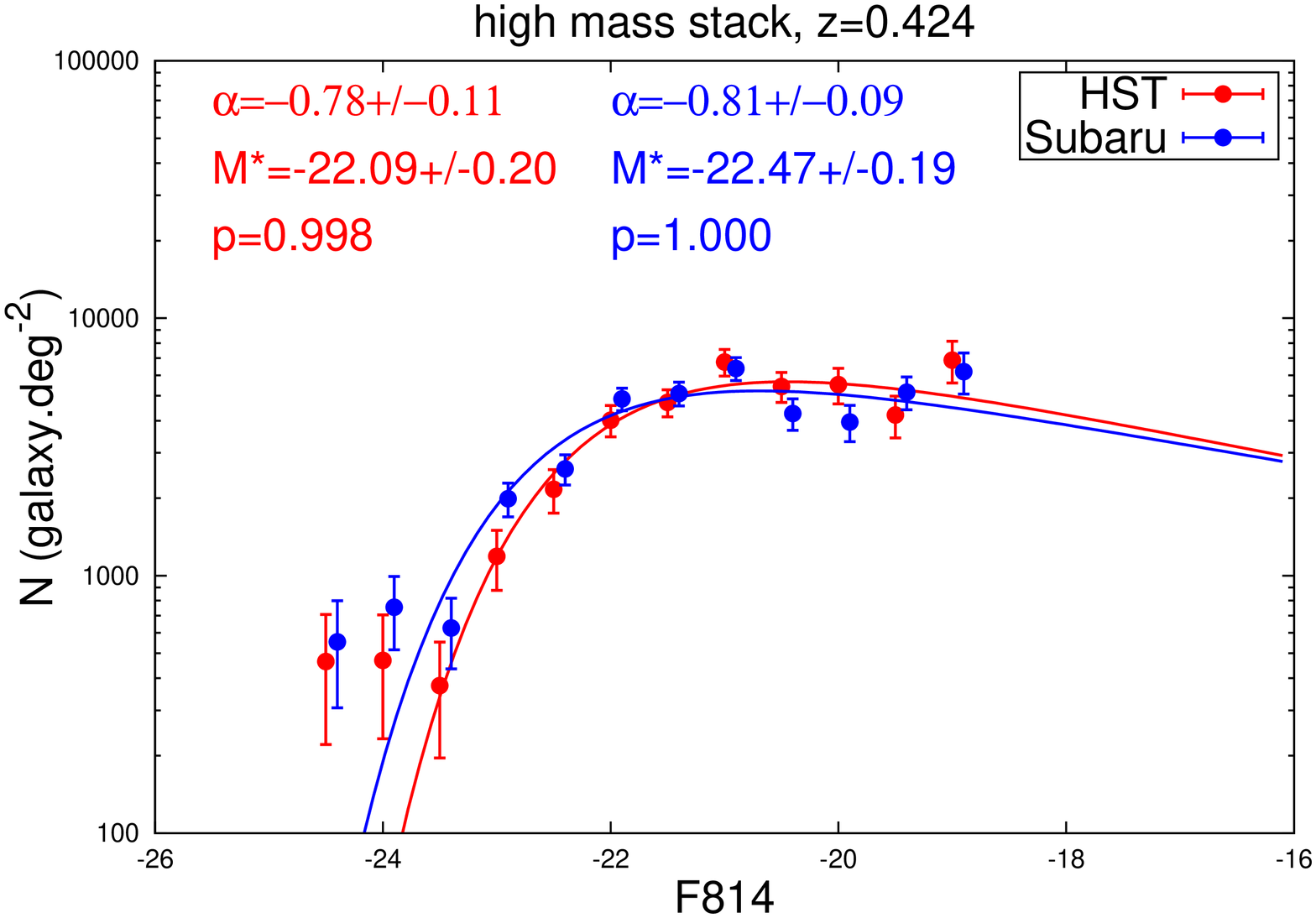}
\caption{Dependence of the stacked cluster GLF on mass, in the F814W
  filter.  {\it Top} represents the low mass 
  ($6\times10^{14}M_\odot<M_{200}<10^{15}M_\odot$) and {\it bottom}
  the high mass ($10^{15}M_\odot<M_{200}$) cluster GLFs. Red and blue correspond to
  the GLFs measured with HST and Subaru respectively, and are
  normalized to one square degree. The curves correspond to the
  Schechter fits to the data. The parameters from each fit are
  displayed in the corresponding color.}
\label{fig:glfstackM}
\end{figure}

Results are shown in Fig.~\ref{fig:glfstackM}. The faint ends from HST
and Subaru agree within the error bars while the characteristic
magnitudes are brighter for Subaru, especially when considering the low
mass sample. We find no significant evolution of the GLF faint end slope with mass, for both
sets of data, with $\alpha=-0.71\pm0.14$ and
$\alpha=-0.78\pm0.11$ for HST low and high mass clusters, and
$\alpha=-0.80\pm0.08$ and $\alpha=-0.81\pm0.09$ for Subaru low and
high mass clusters respectively. However, we note that there might be a
degeneracy with redshift, as the low mass sample has a mean redshift
$\bar{z}=0.360$ and the high mass $\bar{z}=0.424$. If high
mass clusters were showing flatter faint ends, this could compensate
for redshift evolution.

\subsection{Breaking the degeneracy between redshift and mass}
\label{subsec:stackzM}

In this section we try to break the degeneracy between redshift and
mass by making four samples: low mass/low z (4 clusters), low
mass/high z (2 clusters), high mass/low z (5 clusters), and high
mass/high z (5 clusters). Given the few clusters in the low mass/high
z, and the large error bars due to brighter completeness limit in the
high mass/high z sample, we can only investigate the low mass/low z
and high mass/low z samples. The median masses for these two samples are $M_{200}=7.06 \times 10^{14} M_\odot$ and $M_{200}=15.40 \times 10^{14} M_\odot$ for the low and high mass respectively.

\begin{figure}
\centering
 \begin{tabular}{c}
\includegraphics[width=0.34\textwidth,clip,angle=270]{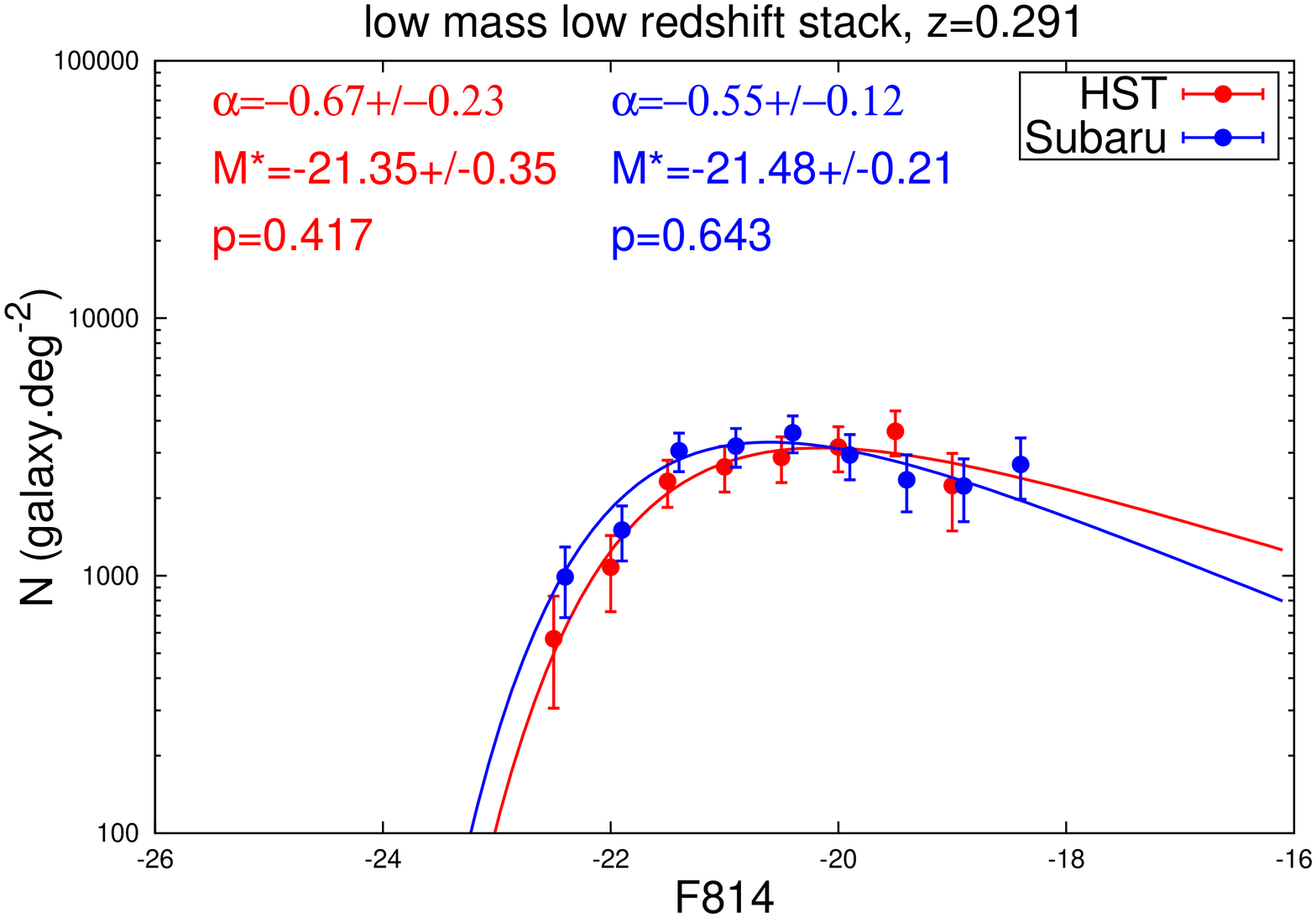}\\
\includegraphics[width=0.34\textwidth,clip,angle=270]{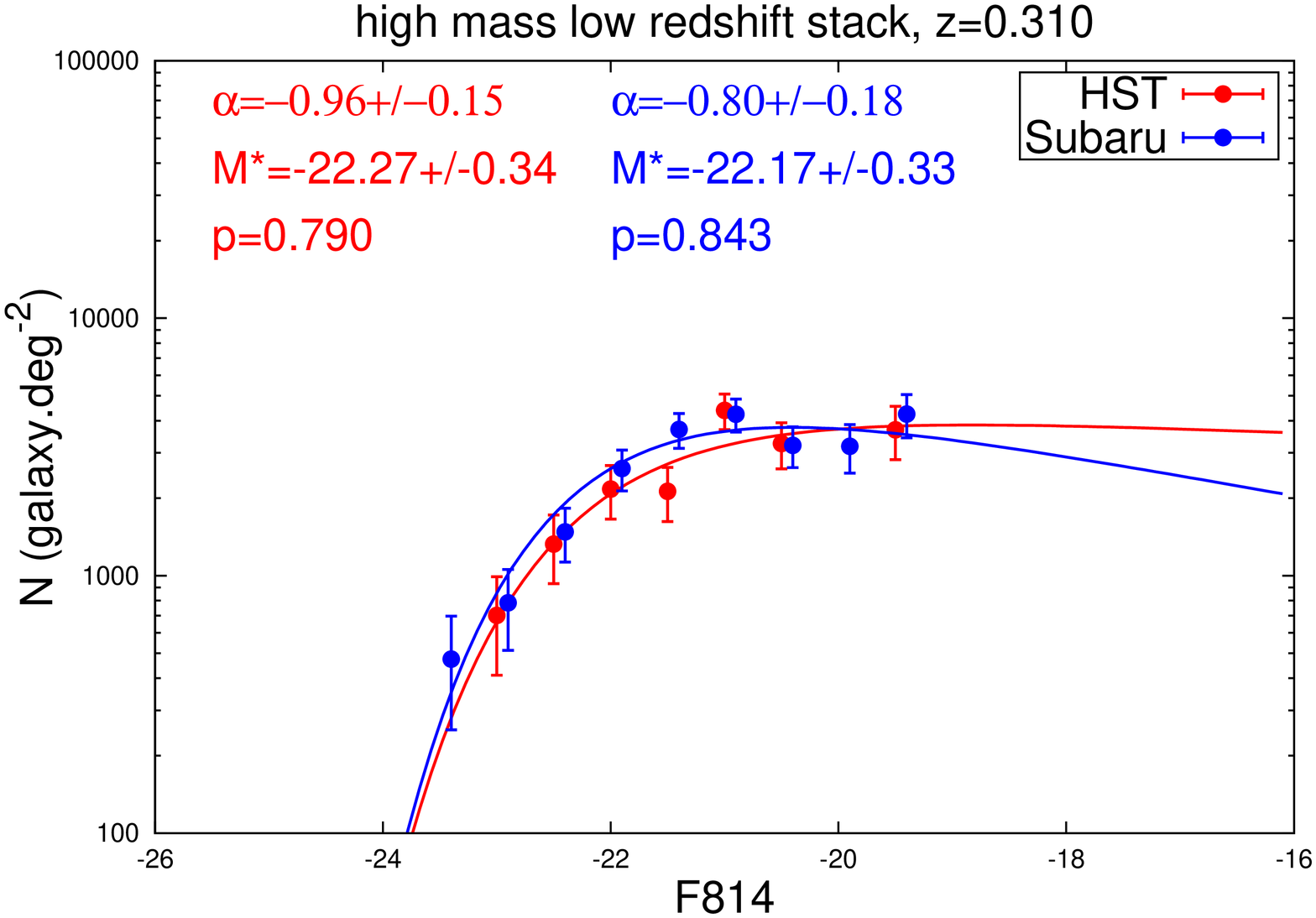}\\
 \end{tabular}
 \caption{Dependence of the stacked cluster GLFs on mass and
   redshift, in the F814W filter. {\it Top} represents the low mass/low
   z GLFs and {\it bottom} the high mass/low z GLFs. Red and blue correspond
   to the GLFs measured with HST and Subaru respectively, and are
   normalized to one square degree. The curves correspond to the
   Schechter fits to the data. The parameters from each fit are
   displayed in the corresponding color.}
\label{fig:glfstackMZ}
\end{figure}

Results are displayed in Fig.~\ref{fig:glfstackMZ}. The HST and
Subaru GLFs agree given the large error bars. We find faint end slopes
$\alpha=-0.67\pm0.23$ and $\alpha=-0.96\pm0.15$ for HST low and high
mass low z samples, and $\alpha=-0.55\pm0.12$ and
$\alpha=-0.80\pm0.18$ for Subaru low and high mass low z samples. This
tends to show that once the degeneracy with redshift is broken, the
high mass clusters show a flatter faint end than the low mass. However,
this hint is detected at only 1.1$\sigma$ and 1.2$\sigma$, so
it would require a larger sample to be verified.

\subsection{Dependence on outer radius}
\label{subsec:rad}

We also take advantage of the large field of view of the Subaru data
to investigate how the GLFs may vary with the radius in which they are
calculated. For this, we first compute individual cluster GLFs in
increasing disks with the following radii: 0.5~Mpc, 1~Mpc, 1.5~Mpc,
2~Mpc, and 2.5~Mpc, before stacking them together. In this way, we
cover different areas from the cluster core to the virial radius.

\begin{figure}
\centering
\includegraphics[width=0.34\textwidth,clip,angle=270]{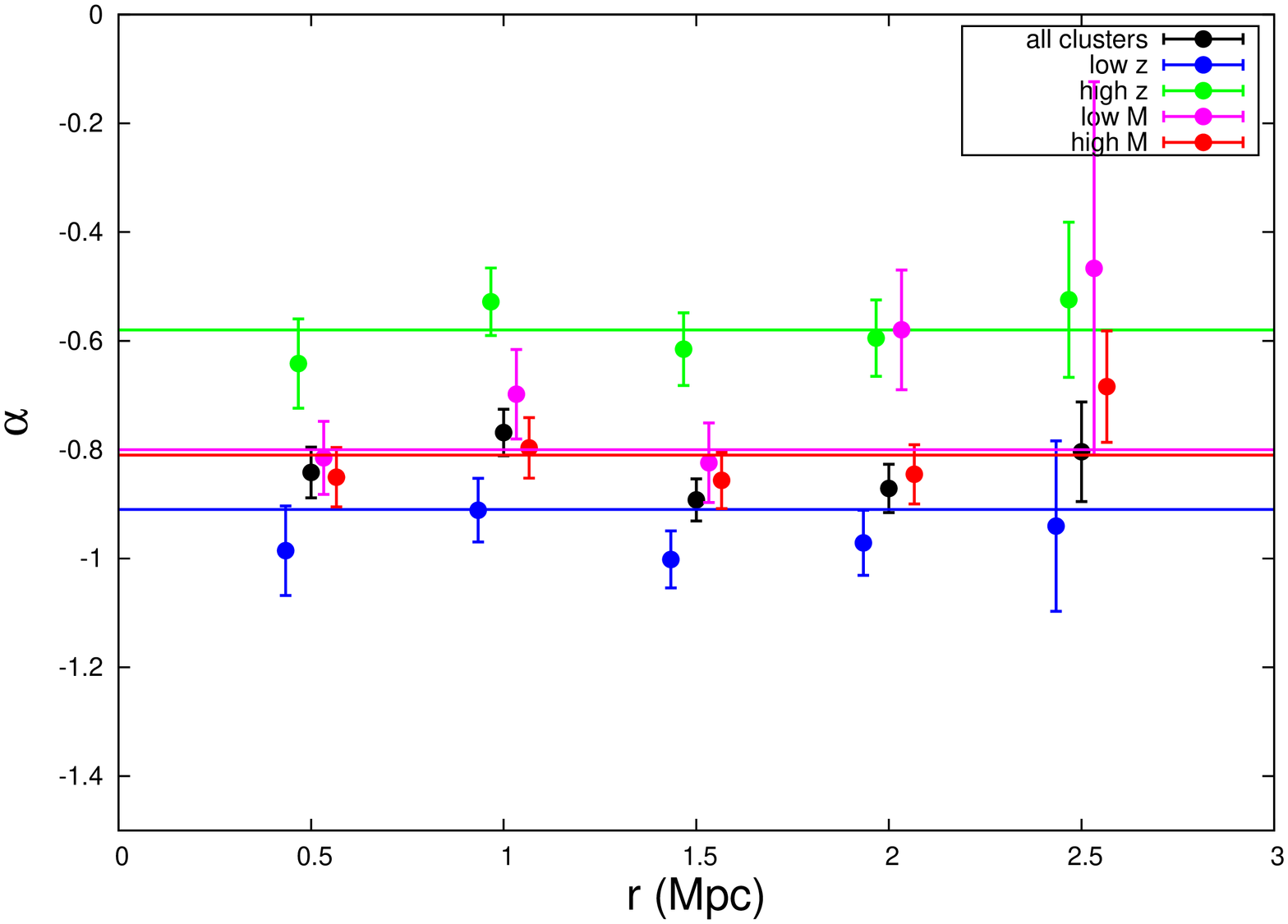}
\includegraphics[width=0.34\textwidth,clip,angle=270]{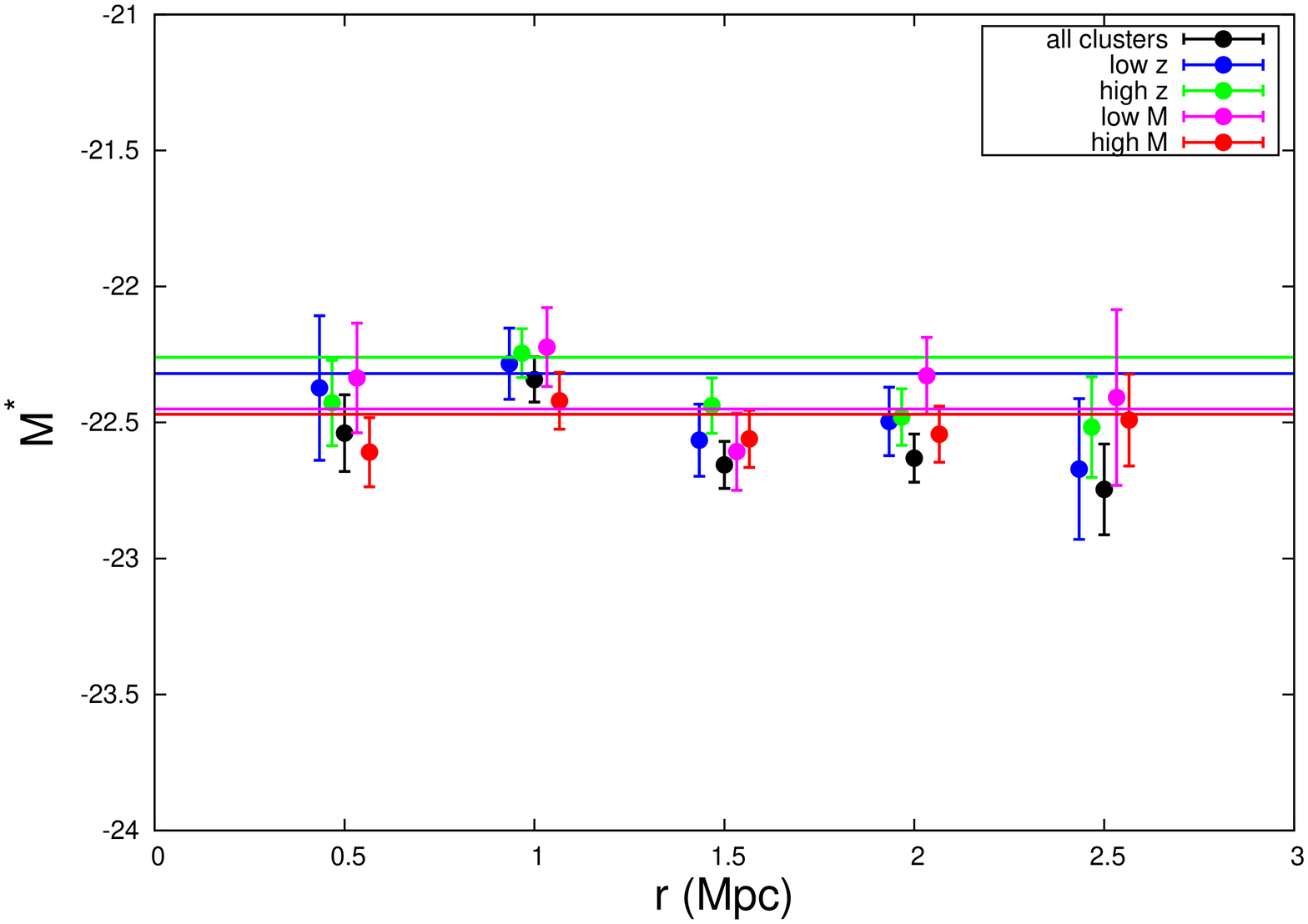}
\caption{Evolution of the stacked Subaru/F814W cluster GLF in disks of
  increasing radius. {\it Top} shows the variation of the $\alpha$ parameter and
  {\it bottom} the variation of the $M^*$ parameter. Every GLF from the stacks is
  computed in a disk centered on the cluster center and extending to
  the radius displayed on the figure. Black dots correspond to
  parameters derived from the stack of all clusters, blue dots
  to low redshift clusters ($0.19<z<0.39$), green dots to
  high redshift clusters ($0.40<z<0.69$), magenta dots
  to low mass clusters ($6\times10^{14}M_\odot<M_{200}<10^{15}M_\odot$), and red dots to
  high mass clusters ($10^{15}M_\odot<M_{200}$). The horizontal lines
  correspond to the values measured from Figs.~\ref{fig:glfstackz} \& ~\ref{fig:glfstackM}. Dots are slightly shifted around their values on the x-axis for clarity.}
\label{fig:glfsubdisk}
\end{figure}

We show in Fig.~\ref{fig:glfsubdisk} the parameters $\alpha$ and $M^*$
as a function of radius for the whole sample (black dots), for the
low (blue) and high (green) redshift samples, and for the low (magenta) and high (red) mass samples, for Subaru
data in the F814W filter. We find no significant variation of the
faint end slope with radius, with values of $\alpha$  consistent with
those found in the previous sections (horizontal lines). The
characteristic magnitudes slightly vary for radii greater than 1~Mpc,
but only at a 1$\sigma$ level. These results
suggest that the GLFs are dominated by the cluster core, and they
don't change when the radius of the considered region increases, at least with the present completeness limit. The only variation seen is in the low mass sample at radii greater than 2~Mpc. These clusters being less rich, they start to be dominated by field galaxies when extending to high radii, leading to less negative faint end slopes $\alpha$ representative of field GLFs \citep[e.g. ][]{Zucca+06,Martinet+15a}. We note however that it does not affect the other results of the paper, as clusters are studied in radii lower than 1~Mpc in the rest of this study due to the limited size of the HST images.

\subsection{Simulating surface brightness dimming}
\label{subsec:SBdim}

In Sect.~\ref{subsec:stackz} we found that the GLF faint end
  shows a mild dependence on redshift in our sample. We now want to
  assess if this redshift evolution can be attributed to the dimming
  of galaxy SB with redshift. Indeed, SB has a dependence on redshift:
  $SB \propto (1+z)^{-4}$. A factor $(1+z)^{-2}$ is due to the dimming
  of the flux when the same galaxy is observed at higher redshift and
  the other $(1+z)^{-2}$ factor accounts for the change in the
    angular area.
  Although we expect to miss some of the higher redshift galaxies
  due to the SB dimming, we note that the dimming is the same for both
  datasets, and is therefore a separate problem from that of the SB
  selection effect between datasets.

If we note $r_a$ the angular radius and $r_p$ the physical radius of a galaxy, those are linked through the angular distance $D_a$:

\begin{equation}
r_a(z)=\frac{r_p}{D_a(z)}=r\frac{1+z}{D_m(z)},
\end{equation}

\noindent where $D_m$ is the comoving distance and $z$ the redshift of the galaxy. Assuming that the physical radius of a galaxy is the same whichever the redshift\footnote{Ignoring size evolution is a conservative estimate as galaxies are smaller at high redshift \citep{vDvdM07} and this would make their SB brighter.}, one can write the fractional change in angular radius from a redshift $z_i$ to $z_f$.

\begin{equation}
\label{eq:3}
\frac{r_a(z_f)}{r_a(z_i)}=\frac{D_m(z_i)}{D_m(z_f)}\frac{1+z_f}{1+z_i}.
\end{equation}

To calculate the change in total flux of the galaxy we not
    only have to consider the change in luminosity distance and the
    k-correction, but also the change in luminosity that comes from
    the luminosity evolution of each galaxy. We adopt the luminosity
    evolution inferred from the fundamental plane evolution, assuming
    that this evolution is purely
  luminosity dependent (i.e. that there is no evolution in the
  physical size). We use equation (10) of \citet{vDvdM07} (recalled in
  eq.~\ref{eq:vFvdM} below) who measured the fundamental plane
  evolution in the redshift range $0.18\leq z \leq 1.28$ which
  includes the redshift interval of the present study.

\begin{equation}
\label{eq:vFvdM}
d\log{(M/L_B)}/dz=-0.555\pm0.042
\end{equation}

\noindent For passive evolution, the mass $M$ of the galaxy
  remains constant and eq.~\ref{eq:vFvdM} allows to calculate
  of the change in luminosity in the B band. We furthermore add a
  k-correction to this relation, noted $k_{B-F814}(z)$, to shift it to
  the F814 filter, and calculate the change in luminosity:

\begin{equation}
\label{eq:lum}
\begin{array}{rcl} 
\log{\left(\frac{L_{F814}(z_f)}{L_{F814}(z_i)}\right)}&=&(-0.555\pm0.042)(z_i-z_f)\\
&&+(k_{B-F814}(z_f)-k_{B-F814}(z_i))/2.5
\end{array}
\end{equation}

\noindent The k-correction is computed using LePhare with the same templates as in the rest of the paper, but adding the B filter to the analysis.

\noindent We can now compute the flux dimming which depends on the luminosity distance $D_l$ and on the luminosity ratio of eq.~\ref{eq:lum}. The conversion from bolometric flux to the given filters is taken into account in the k-correction in the luminosity term and therefore does not appear in this equation.

\begin{equation}
\label{eq:6}
\begin{array}{rcl} 
\frac{F_{F814}(z_f)}{F_{F814}(z_i)}&=&\frac{L_{F814}(z_f)}{L_{F814}(z_i)}\left(\frac{D_l(z_i)}{D_l(z_f)}\right)^2\\&=&\frac{L_{F814}(z_f)}{L_{F814}(z_i)}\left(\frac{(1+z_i)D_m(z_i)}{(1+z_f)D_m(z_f)}\right)^2
\end{array}
\end{equation}

We can use the set of equations given above to compute the SB
  dimming of cluster galaxies when we shift them from low redshift to
  high redshift. In particular we want to simulate images of the low
  redshift cluster sample as they would appear at higher
  redshift. We shift every low redshift cluster by the difference in
  mean redshift of the high and low redshift samples. Comparing
  the stacked GLFs of the low and high redshift simulated images to
  those of the observed data allows to check whether SB
  dimming can explain the observed redshift evolution or not.

Simulations are made with the GalSim software \citep{galsim},
  using galaxies measured on the observed images. The PSF is measured
  on each image with PSFEx, and Sersic profiles convolved with this
  PSF are fitted to galaxies using SExtractor. The parameters from this
  fit are then used in GalSim to simulate galaxies as single Sersic
  profiles with half-light radius, Sersic index, flux, and position
  from the original image. In the case of the high-redshift simulations, we apply the evolution given in eqs.~\ref{eq:3} \& \ref{eq:6} to the radii and fluxes measured on the low-redshift galaxies before using these quantities in the simulations. These profiles are then convolved with an
  analytic Moffat PSF with $\beta = 4.765$ \citep[following prescriptions from][]{Trujillo+01} and the measured
  FWHM, before being inserted into the images. The pixel scale is the same as
  in the data (0.03'' for HST and 0.2'' for Subaru) and we add
  Gaussian random noise with the sky rms value measured in the image
  by SExtractor. We keep the same noise seed for the fiducial
  simulations (with the low redshift clusters) and the high redshift ones.\\

\begin{figure}
\centering
\includegraphics[width=0.34\textwidth,clip,angle=270]{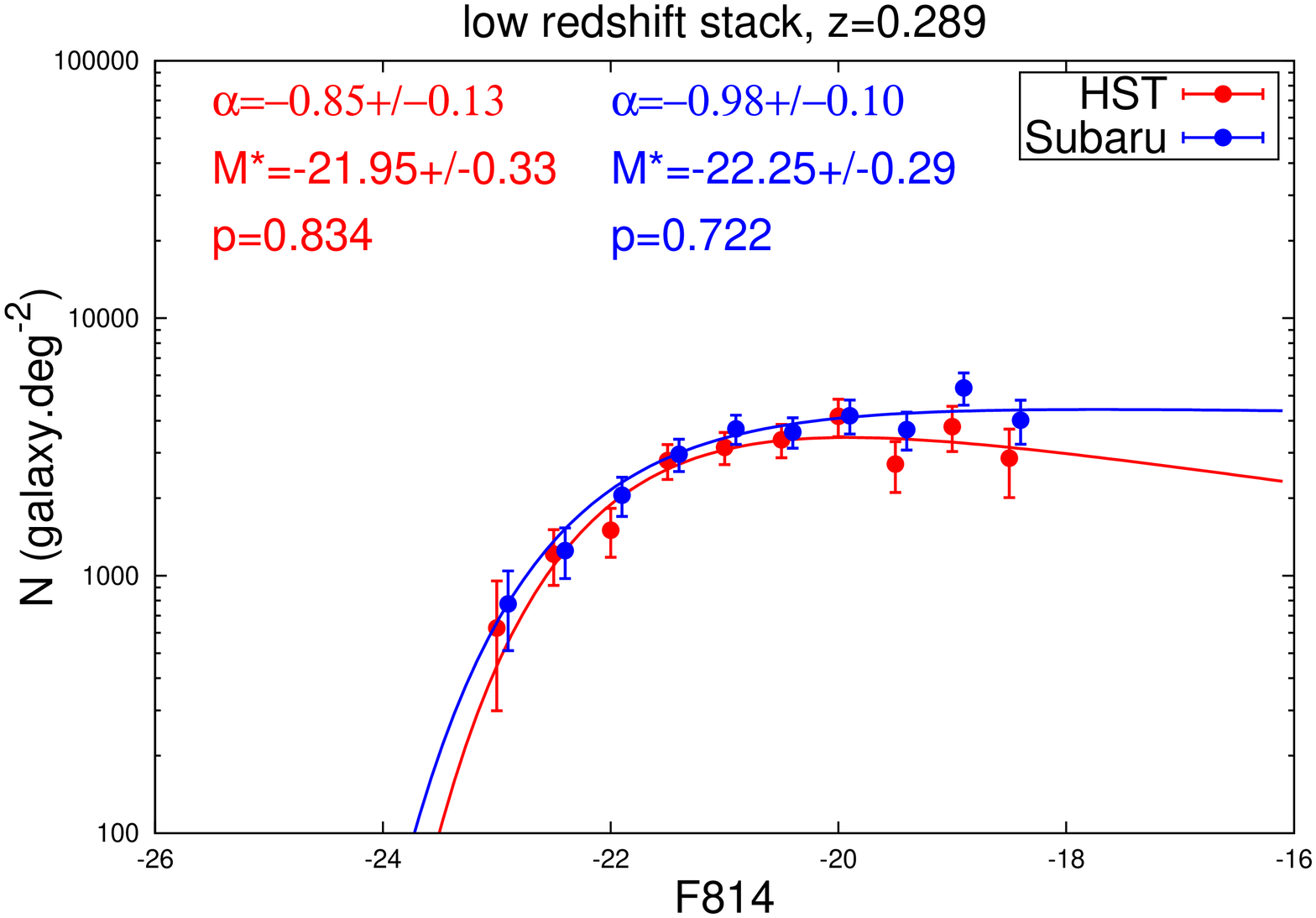}
\includegraphics[width=0.34\textwidth,clip,angle=270]{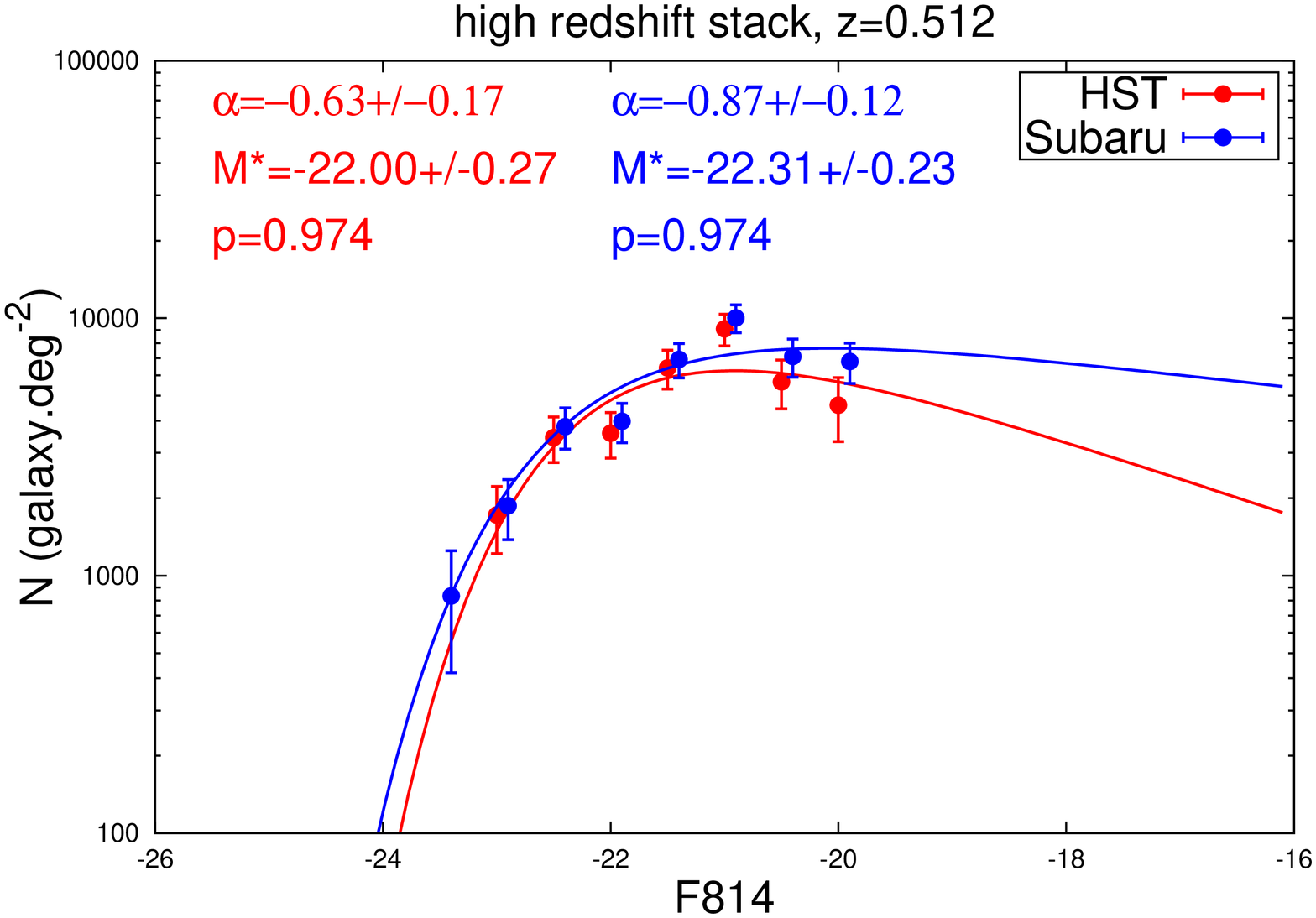}
\caption{Effect of SB dimming when evolving the low
  redshift sample to high redshift. {\it Top} represents the low
  redshift simulated stack GLF ($\bar{z}=0.289$) and {\it bottom} the
  GLF of the same clusters simulated at high redshift
  ($\bar{z}=0.511$). Red and blue correspond to the GLFs measured with
  HST and Subaru respectively, and are normalized to one square
  degree. The curves correspond to the Schechter fits to the data. The
  parameters from each fit are displayed in the corresponding color.}
\label{fig:glfstackdim}
\end{figure}

Figure~\ref{fig:glfstackdim} shows the stacked GLFs for the low
  redshift simulated sample and the same clusters evolved to the mean
  redshift of the high redshift sample. These GLFs are to be compared
  with the observed redshift evolution
  (Fig.~\ref{fig:glfstackz}). First we see that the simulated
  low-redshift GLF is not identical to the observed one due to the
  fact that the simulations are simplistic compared to real
  data. However the simulated and observed GLFs agree within the error
  bars with a difference of 0.6$\sigma$ for HST and 0.5$\sigma$ for
  Subaru, validating our simulation pipeline.

Looking at the simulated high-redshift sample we find that the
  slope of the GLF has slightly decreased compared to the low redshift
  simulation. We find that the slope $\alpha$ for the simulated
    clusters varies by 1.0$\sigma$ and 0.7$\sigma$ from low to high
  redshift respectively for HST and Subaru. We interpret this
    change as coming from SB selection. These values are to be
  compared with the 1.7$\sigma$ and 2.6$\sigma$ variation in the case
  of the observations. We therefore conclude that SB dimming is not
  sufficient to explain alone the observed redshift evolution of the
  faint end of the GLF. The small observed evolution may also be due to our relatively small redshift baseline, and a larger range of redshifts would be valuable to secure our findings. We also investigate how robust these results
  are to the knowledge of the fundamental plane by computing the
  variation of the faint end slope when applying the error bars of
  eq.~\ref{eq:vFvdM}. The slope $\alpha$ of the evolved stacked GLF
  varies of $\pm0.6\sigma$ and $\pm0.4\sigma$, in the case of HST and
  Subaru respectively, when considering these error bars. 
  We note that $M^*$ is not significantly affected by SB dimming,
  which we can expect as the bright galaxies should be visible
    regardless of the SB dimming.

\section{Discussion}
\label{sec:discussion}

\begin{figure}
\centering
\includegraphics[width=0.34\textwidth,clip,angle=270]{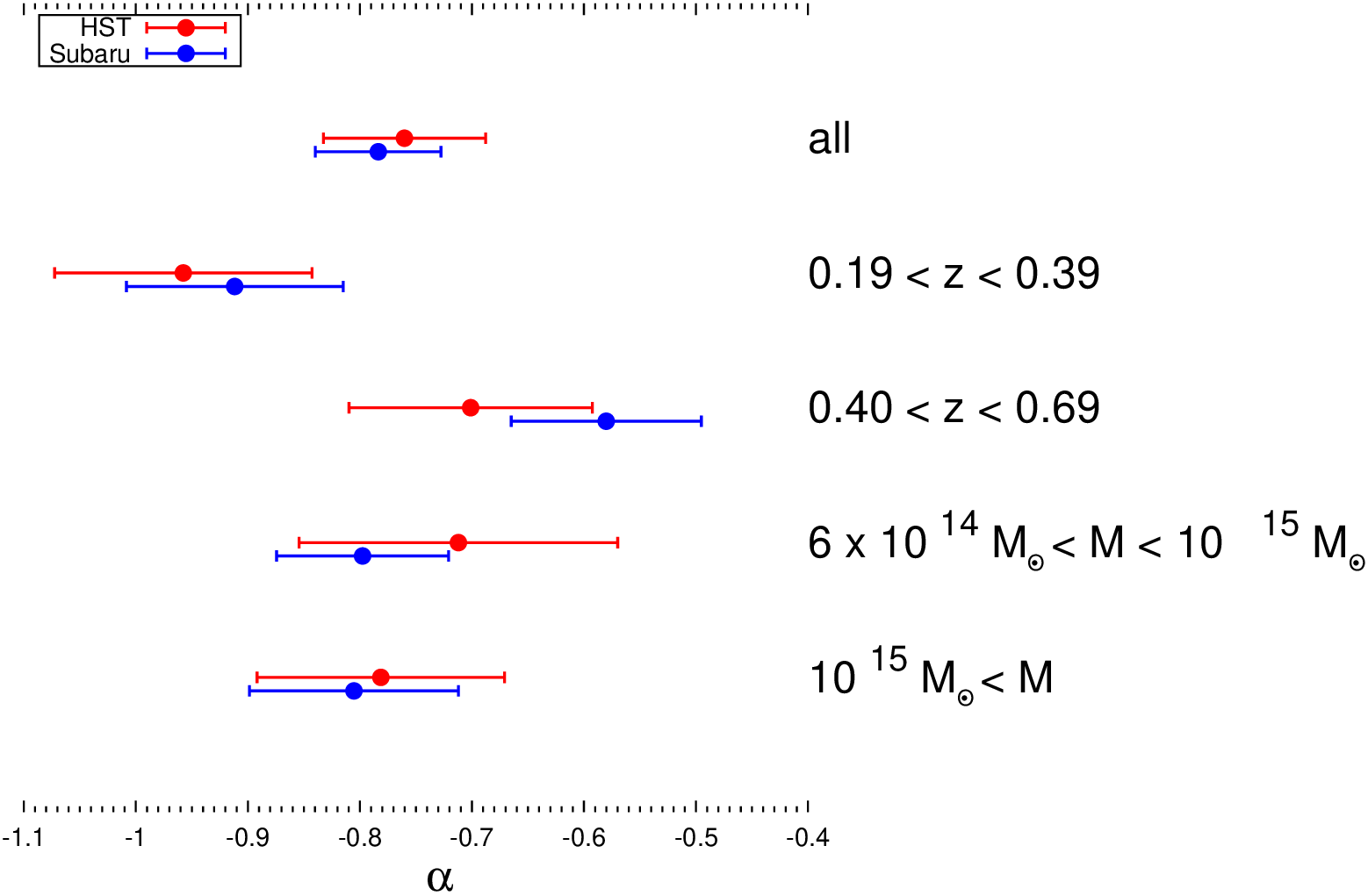}
\includegraphics[width=0.34\textwidth,clip,angle=270]{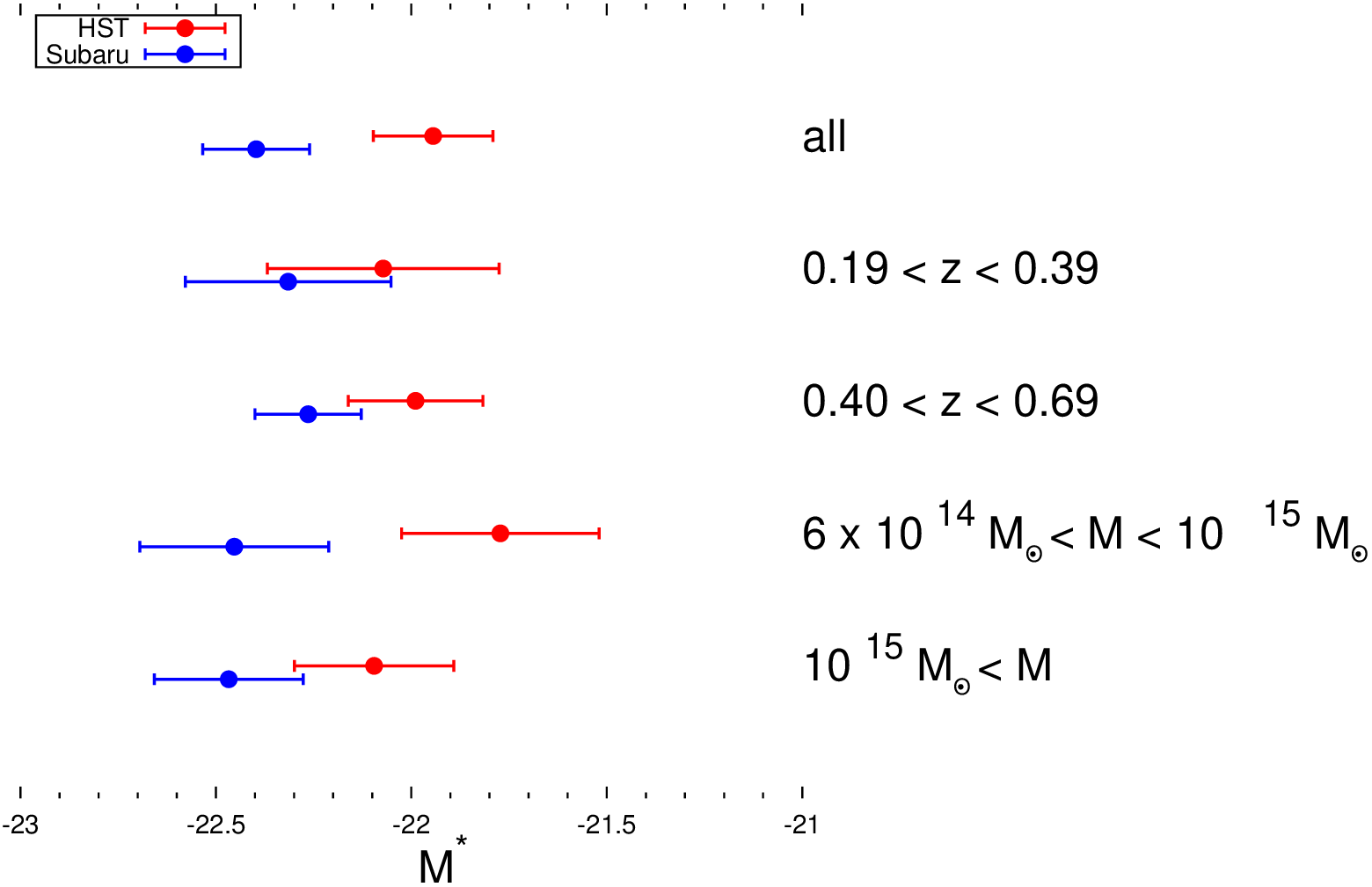}
\caption{Evolution of $\alpha$ and $M^*$ with redshift and mass. Red and blue
   correspond to the GLFs measured with HST and Subaru, in the F814W filter.}
\label{fig:glffinal}
\end{figure}

We compute the stacked GLFs for 16 CLASH clusters based on independent
HST and Subaru analyses in order to study the faint end and the
characteristic magnitude of cluster RS GLFs in the redshift range
$0.187\leq z\leq0.686$, and their evolution with redshift and mass. A
summary of the main results can be found in Fig.~\ref{fig:glffinal},
where we plot $\alpha$ and $M^*$ values for the whole sample and for the
subsamples in the different redshift and mass ranges. The use of both
space and ground based data enables us to discuss selection effects, while simulations permit to investigate the effect of SB dimming with redshift (cf. Fig.~\ref{fig:glfstackdim}). In addition, the Subaru large field of view allows
to study the dependence on radius, and results on this point are
displayed in Fig.~\ref{fig:glfsubdisk} in
Sect.~\ref{subsec:rad}. Based on these figures, the main results
of our analysis are the following:\\

1. We find no dependence of $\alpha$ or $M^*$ on radius in the range
0.5 to 2.5~Mpc, except for low mass clusters at radius greater than 2~Mpc. This probably means that cluster GLFs are dominated by
the cluster core in the probed magnitude range. We can therefore
be certain that the smaller size of the HST field of view is not
responsible for the different faint end behaviors observed in the
literature.\\

2. We find no evolution of $M^*$ either with redshift or mass,
suggesting that the bright population is similar in the studied
redshift and mass ranges. We recall that the lowest mass of our
clusters is $6 \times 10^{14}M_\odot$, so they are all quite massive,
explaining that they have similar abundances of bright
galaxies. However, we find that the $M^*$ value derived from HST is
$\sim0.4$ magnitude fainter than that from Subaru. A possible
explanation would be a leakage of stars into the GLF bright end, as
the star-galaxy separation is not as good in the Subaru data due to
the larger PSF. However, we could not find evidence for this by visual
inspection, as it is difficult to discriminate between
stars and galaxies with circular shapes at these magnitudes. Another explanation might reside in the low statistics of the background subtraction at the bright end, which could introduce small differences between the bright ends of the two data sets.\\

3. Using the whole sample ($\bar{z}=0.40$), we find a decreasing
  faint end for both datasets with consistent values between HST
  ($\alpha=-0.76\pm0.07$) and Subaru
  ($\alpha=-0.78\pm0.06$). Separating between a low redshift
  ($\bar{z}=0.29$) and high redshift ($\bar{z}=0.51$) samples, we find
  an evolution of the faint end slope of 1.7$\sigma$ with HST and
  2.6$\sigma$ with Subaru. There is thus a mild decrease of the faint
  end slope (less negative $\alpha$) with increasing redshift over the
  range ($0.187<z<0.686$). This evolution is in good agreement with
  recent papers in the literature: in particular \citet{Zenteno+16}
  found a decrease of the RS faint end at $2.1\sigma$ for a wider
  range of redshifts ($0.1<z<1.13$), but with $\sim80\%$ of their
  clusters being in the same redshift range as ours.

  \citet{DePropris+13} claimed that the evolution in the faint end
  slope has a significant contribution from surface brightness
  selection effects.  They used HST data of differing depths on a
  single cluster (MS~1358.4+6254) to show that surface brightness
  selection effects become important above the formal magnitude limit
  of their data and that they affect the red sequence GLF at
  magnitudes z$\geq$24.5 for 2.7~ksec HST exposures (see their
  Fig.~18).  Given that the faint red sequence for their cluster has
  $F814W-z=0.25$, this implies that the SB selection effects in their
  sample become important at $F814W>24.75$.  On the other hand, our
  CLASH data are significantly deeper than theirs (4.1~ksec) and we
  limit our GLFs at $F814W<24.5$. Therefore, the real SB selection
  effects noticed in \citet{DePropris+13} should not be playing a role
  in our space-based results.

  In addition, \citet{DePropris+13} claim that previous estimates of
  the evolution in the red sequence GLF
  \citep[e.g. ][]{DeLucia+07,Rudnick+09} were also due to SB effects.
  Both of those works were based on the same ground-based data with a
  formal magnitude limit of I$=24$ or 24.5 (for the low and high redshift
  clusters respectively) and the evolution in the GLF was seen over
  the faintest 2 magnitudes.  We cannot directly address the role of
  SB effects in the EDisCS results without detailed simulations on
  those data (see below for such simulations for our clusters) but the
  similarity between our HST and Subaru GLFs imply that the EDisCS
  evolution in the GLF is not dominated by SB effects.\\

  4. We artificially evolved the low redshift clusters to high
    redshifts, through simulations taking into account the fundamental
    plane evolution and SB dimming. Computing the GLFs
    from these simulations we find no evolution of $M^*$ with redshift
    and no significant evolution of $\alpha$, namely $1.0\sigma$ and
    $0.7\sigma$ for HST and Subaru respectively. Surface brightness dimming
    therefore cannot explain the redshift evolution of the GLF
    observed in the data.\\

5. We see no significant trend of the faint end with mass, but
maybe because all of our clusters are quite massive. We note that this
result agrees with the weak dependence on mass found in
e.g. \citet{DeLucia+07, Gilbank+08, Rudnick+09}. \citet{Cerulo+16}
found hints that more massive clusters could have flatter GLFs for
high redshift clusters ($0.8<z<1.5$), but it seems not to be the case
at lower redshifts, at least in the mass range probed here.\\

Though SB selection effects and SB dimming affect the high
redshift cluster GLFs, they are not sufficient to explain alone the deficit of
RS galaxies up to $z\sim0.7$, which therefore requires some physical process such as quenching of star formation. As there is also no dependence of the
GLFs on the image field of view, the number of possible explanations to the
differences found in the literature becomes smaller. One last point worth
investigating is the selection of clusters, as all studies, including
the present one, select small sets of clusters (typically a few to a few
tens of clusters), and are based on different criteria. The only way to
uncover this problem is to build a very large sample of galaxy clusters, such
as the one that will be available in upcoming large optical surveys.

\begin{acknowledgements}

  We thank the referee for useful comments which improved the quality and readability of the paper. We are very grateful to the CLASH team for making their data
    publicly available. For the same reason we also thank the 3D-HST
    team for the COSMOS reduced images. Finally we thank the GalSim
    team and Emmanuel Bertin for their respective public software and
    additionally Emmanuel for useful discussions. Florence Durret
  acknowledges long-term support from CNES. Gregory Rudnick
    acknowledges support of NSF funding from proposals 1517815 and
    1211358.

\end{acknowledgements}

\bibliographystyle{aa}
\bibliography{glf}

\begin{appendix}

\section{Individual cluster GLFs: Schechter parameters}
\label{appendix:glfparams}

The Schechter parameters for the individual cluster GLFs are given in 
Table~A.1.

\begin{landscape}
\begin{table}
\centering
\begin{tabular}{lc|ccccc|ccccc}
\hline
\hline
&&& & F606W&&  && & F814W&& \\
\hline
&$z$&$\alpha$ & M*  &$\phi$* ($deg^{-2}$)&                comp & $p$ &                                                      $\alpha$ & M*  &$\phi$* ($deg^{-2}$) &                            comp & $p$ \\
\hline
A383 & 0.19&&&& & & & & & & \\
HST & & -1.12$\pm$    0.17 & -21.5$\pm$  0.7 &  2176$\pm$  1306 &  -16.2 &  0.787 & -1.25$\pm$    0.12 & -24.4$\pm$  1.9 &   706$\pm$   674 &  -17.4 &  0.587 \\
Subaru & & -0.90$\pm$    0.19 & -21.3$\pm$  0.6 &  4073$\pm$  1886 &  -16.2 &  0.971 & -1.16$\pm$    0.13 & -23.6$\pm$  0.9 &  1474$\pm$   899 &  -17.4 &  0.591 \\
\hline
A209 & 0.21&&&& & & & & & & \\
HST & & -1.22$\pm$    0.12 & -22.7$\pm$  1.1 &  1672$\pm$  1219 &  -16.0 &  0.956 & -1.22$\pm$    0.10 & -23.5$\pm$  1.1 &  1461$\pm$   933 &  -16.8 &  0.930 \\
Subaru & & -1.29$\pm$    0.09 & -23.4$\pm$  1.1 &  1126$\pm$   775 &  -16.0 &  0.769 & -1.23$\pm$    0.12 & -22.9$\pm$  0.8 &  1738$\pm$  1085 &  -16.7 &  0.940 \\
\hline
RXJ2129 & 0.23&&&& & & & & & & \\
HST & & -0.70$\pm$    0.50 & -20.7$\pm$  0.8 &  5706$\pm$  3897 &  -17.9 &  0.751 & -0.75$\pm$    0.48 & -21.2$\pm$  0.8 &  5504$\pm$  3880 &  -18.1 &  0.819 \\
Subaru & & -1.05$\pm$    0.35 & -21.3$\pm$  0.9 &  3651$\pm$  3240 &  -17.9 &  0.969 & -0.98$\pm$    0.38 & -21.8$\pm$  0.9 &  3894$\pm$  3406 &  -18.1 &  0.976 \\
\hline
A611 & 0.29&&&& & & & & & & \\
HST & & -0.06$\pm$    1.05 & -20.4$\pm$  0.9 & 13874$\pm$  4078 &  -19.0 &  0.640 &  0.03$\pm$    0.71 & -20.8$\pm$  0.6 & 14141$\pm$  3173 &  -19.1 &  0.622 \\
Subaru & &  0.63$\pm$    1.88 & -20.0$\pm$  1.0 & 12367$\pm$  8026 &  -19.0 &  0.834 &  0.27$\pm$    1.42 & -20.9$\pm$  0.9 & 12888$\pm$  2328 &  -19.1 &  0.991 \\
\hline
MS2137 & 0.31&&&& & & & & & & \\
HST & & -1.81$\pm$    2.56 & -22.0$\pm$  4.2 &  1147$\pm$  8555 &  -20.3 &  0.182 & -0.65$\pm$    1.26 & -21.7$\pm$  1.3 &  3450$\pm$  3608 &  -20.3 &  0.525 \\
Subaru & & -1.38$\pm$    1.32 & -21.8$\pm$  2.0 &  3109$\pm$  8242 &  -20.3 &  0.064 & -0.15$\pm$    0.79 & -21.4$\pm$  0.6 &  7942$\pm$  2276 &  -20.3 &  0.388 \\
\hline
RXJ1532 & 0.34&&&& & & & & & & \\
HST & & -0.33$\pm$    0.66 & -20.5$\pm$  0.7 & 10413$\pm$  4063 &  -17.7 &  0.502 & -0.30$\pm$    0.49 & -21.0$\pm$  0.6 & 10409$\pm$  3365 &  -18.1 &  0.406 \\
Subaru & & -0.37$\pm$    0.42 & -20.6$\pm$  0.7 & 10024$\pm$  4002 &  -17.7 &  0.960 & -0.16$\pm$    0.26 & -21.2$\pm$  0.3 & 12004$\pm$  2017 &  -18.1 &  0.787 \\
\hline
MACSJ1115 & 0.35&&&& & & & & & & \\
HST & & -0.41$\pm$    0.36 & -21.1$\pm$  0.4 & 11353$\pm$  3508 &  -18.1 &  0.984 & -0.57$\pm$    0.22 & -21.5$\pm$  0.4 & 12807$\pm$  3714 &  -17.6 &  0.551 \\
Subaru & & -0.59$\pm$    0.26 & -21.1$\pm$  0.3 & 10480$\pm$  3211 &  -18.1 &  0.799 & -0.56$\pm$    0.21 & -21.9$\pm$  0.3 & 10336$\pm$  2977 &  -17.6 &  0.898 \\
\hline
MACSJ1720 & 0.39&&&& & & & & & & \\
HST & & -0.80$\pm$    0.35 & -22.0$\pm$  0.6 &  7384$\pm$  4198 &  -18.6 &  0.916 & -0.78$\pm$    0.33 & -22.3$\pm$  0.5 &  7173$\pm$  3573 &  -19.4 &  0.977 \\
Subaru & & -1.23$\pm$    0.30 & -22.4$\pm$  1.0 &  3225$\pm$  3216 &  -18.6 &  0.624 & -1.04$\pm$    0.22 & -23.1$\pm$  0.7 &  4552$\pm$  2796 &  -19.4 &  0.707 \\
\hline
MACSJ0429 & 0.40&&&& & & & & & & \\
HST & & - & - &     - &  -19.2 &  - & -0.93$\pm$    0.29 & -21.4$\pm$  0.7 &  6864$\pm$  4177 &  -18.0 &  0.783 \\
Subaru & &  0.40$\pm$    0.80 & -20.1$\pm$  0.5 & 12688$\pm$  2863 &  -19.2 &  0.860 & -0.13$\pm$    0.42 & -21.2$\pm$  0.5 & 11973$\pm$  2930 &  -18.0 &  0.924 \\
\hline
MACSJ1206 & 0.44&&&& & & & & & & \\
HST & & -0.64$\pm$    0.17 & -21.4$\pm$  0.3 & 19228$\pm$  4481 &  -18.4 &  0.999 & -0.64$\pm$    0.17 & -21.9$\pm$  0.3 & 18184$\pm$  4269 &  -18.8 &  0.999 \\
Subaru & & -0.70$\pm$    0.15 & -21.9$\pm$  0.3 & 15520$\pm$  3896 &  -18.4 &  0.906 & -0.58$\pm$    0.17 & -22.2$\pm$  0.3 & 18173$\pm$  4074 &  -18.8 &  0.995 \\
\hline
MACSJ0329 & 0.45&&&& & & & & & & \\
HST & & -0.66$\pm$    0.29 & -21.5$\pm$  0.3 & 17008$\pm$  5178 &  -19.1 &  0.990 & -0.54$\pm$    0.23 & -21.8$\pm$  0.3 & 18812$\pm$  4508 &  -19.4 &  0.927 \\
Subaru & & -0.55$\pm$    0.29 & -21.6$\pm$  0.4 & 16801$\pm$  4957 &  -19.1 &  0.942 & -0.47$\pm$    0.22 & -22.2$\pm$  0.3 & 17184$\pm$  3932 &  -19.4 &  0.962 \\
\hline
RXJ1347 & 0.45&&&& & & & & & & \\
HST & &  0.19$\pm$    0.62 & -20.4$\pm$  0.4 & 18248$\pm$  2539 &  -18.6 &  0.996 &  0.14$\pm$    0.48 & -20.9$\pm$  0.4 & 17931$\pm$  2431 &  -18.4 &  0.995 \\
Subaru & & -0.01$\pm$    0.45 & -20.6$\pm$  0.4 & 18032$\pm$  3032 &  -18.6 &  0.971 & -0.52$\pm$    0.13 & -21.9$\pm$  0.2 & 13436$\pm$  2238 &  -18.4 &  0.954 \\
\hline
MACSJ1423 & 0.55&&&& & & & & & & \\
HST & & -0.59$\pm$    0.42 & -21.7$\pm$  0.6 &  9613$\pm$  4589 &  -19.5 &  0.564 & -0.47$\pm$    0.32 & -21.9$\pm$  0.4 & 10500$\pm$  3341 &  -19.0 &  0.766 \\
Subaru & & -0.59$\pm$    0.34 & -21.6$\pm$  0.5 & 10491$\pm$  3959 &  -19.5 &  0.638 & -0.67$\pm$    0.18 & -22.6$\pm$  0.3 &  9187$\pm$  2760 &  -19.0 &  0.924 \\
\hline
MACSJ0717 & 0.55&&&& & & & & & & \\
HST & & -1.11$\pm$    0.51 & -22.5$\pm$  0.7 & 11048$\pm$  9466 &  -21.0 &  0.783 & -0.84$\pm$    0.44 & -22.5$\pm$  0.6 & 15871$\pm$  9138 &  -20.5 &  0.836 \\
Subaru & & -0.73$\pm$    0.49 & -21.8$\pm$  0.4 & 18568$\pm$  7307 &  -21.0 &  0.997 & -0.62$\pm$    0.28 & -22.7$\pm$  0.4 & 16367$\pm$  4885 &  -20.5 &  0.987 \\
\hline
MACSJ2129 & 0.57&&&& & & & & & & \\
HST & & -0.43$\pm$    1.13 & -21.2$\pm$  1.0 & 17761$\pm$ 10170 &  -20.2 &  0.618 & -0.55$\pm$    0.31 & -21.7$\pm$  0.4 & 15777$\pm$  5236 &  -19.1 &  0.910 \\
Subaru & & -0.52$\pm$    0.45 & -21.4$\pm$  0.4 & 19823$\pm$  5473 &  -20.2 &  0.337 & -0.37$\pm$    0.16 & -22.0$\pm$  0.2 & 18698$\pm$  2794 &  -19.1 &  0.985 \\
\hline
MACSJ0744 & 0.69&&&& & & & & & & \\
HST & & - & - &   - &  -21.5 &  - & -2.00$\pm$    0.41 & -24.9$\pm$  2.7 &   371$\pm$  1416 &  -21.3 &  0.013 \\
Subaru & & - & - &   - &  -21.5 &  - & -1.59$\pm$    0.31 & -25.6$\pm$  2.9 &   361$\pm$   960 &  -21.3 &  0.994 \\
\hline
\hline
\end{tabular}
\caption{Parameters from the Schechter fit to cluster RS GLFs for HST and Subaru. ''comp'' and ''p'' correspond to the completeness limit to which the fit is done and the goodness of the fit normalized to 1. '-' indicates that the fit did not converge.}
\end{table}
\end{landscape}

\end{appendix}

\end{document}